\documentstyle[preprint,aps,epsf]{revtex}

\def\by#1#2{{\displaystyle {#1}\over \displaystyle {#2}}}
\def\d{{\rm d}}
\preprint {IMSc/2004/07/30}
\def\by#1#2{{\displaystyle {#1}\over \displaystyle {#2}}}
\def\d{{\rm d}}

\begin{document}
\title{A question of hierarchy: matter effects with atmospheric neutrinos 
and anti-neutrinos} \author{D. Indumathi and  M. V. N. Murthy}

\address
{The Institute of Mathematical Sciences, Chennai 600 113, India.\\
}
\date{\today}
\maketitle
\begin{abstract}

It is by now established that neutrinos mix, have (different) non-zero
masses, and therefore oscillate. The oscillation parameters themselves,
however, are not all well-known. An open problem is that of the neutrino
mass hierarchy. We study the possibility of determining the neutrino
mass hierarchy with atmospheric neutrinos using an iron calorimeter detector
capable of charge identification such as the proposed {\sc monolith} and
{\sc ical/ino} detectors. We find that such detectors are sensitive to
the sign of the mass-squared difference, $\delta_{32} = m_3^2 - m_2^2$,
provided the as-yet unknown mixing angle between the first and third
generations, $\theta_{13}$, is greater than $6^\circ$ ($\sin^2 2\theta_{13}
> 0.04$). A result with a significance greater than 90\% CL requires large
exposures (more than 500 kton-years) as well as good energy and angular
resolution of the detected muons (better than 15\%), especially for small
$\theta_{13}$. Hence obtaining definitive results with such a detector
is difficult, unless $\theta_{13}$ turns out to be large. In contrast,
such detectors can establish a clear oscillation pattern in atmospheric
neutrinos in about 150 kton-years, therefore determining the absolute
value of $\delta_{32}$ and $\sin^2 2 \theta_{23}$ to within 10\%.

\end{abstract}

\pacs{PACS numbers:}

\narrowtext

\section{Introduction}

Recent experimental results from the observations of solar, atmospheric
and reactor neutrinos have provided compelling evidence for neutrino
oscillations \cite{solar,atm,reactor}. Until recently the evidence for
neutrino oscillation was indirect, with a small window still available
for other possibilities such as neutrino decay. The latest results from
the Super-Kamiokande (Super-K) collaboration \cite{ishi}, presenting the
evidence for an ``oscillation minimum" in the observation of atmospheric
neutrinos, seem to provide for the first time clear and unambiguous
evidence for the oscillation hypothesis \cite{theory,mixing} and hence
of non-zero (and different) neutrino masses and mixing. Recently, data
is also available from the long-baseline experiment K2K\cite{ktok}.

Clearly the focus from now on will be to determine precisely the mixing
parameters and mass squared differences apart from the observation of
the oscillation pattern beyond the first minimum. An important question
that still remains open is the hierarchy of the masses. Within the three
neutrino mixing framework, the results on solar neutrinos prefer the
dominant mass eigenstates in $\nu_e$ to have the hierarchy $m_2 > m_1$ so
that the mass-squared difference\footnote{All oscillation experiments are
sensitive to mass-squared differences rather than the absolute scale of
the masses themselves.} $\delta_{21} = m_2^2 - m_1^2 > 0$. The sign of the
other mass-squared difference $\delta_{32}$ (equivalently $\delta_{31}$
since it is so much larger than $\delta_{21}$) is not known.

The issue of the neutrino mass hierarchy (and measurement of the (13)
mixing angle) was discussed by Tabarelli de Fatis \cite{tabarelli}
through a measurement of muon neutrino and anti-neutrino events in a
magnetised iron calorimeter.  In a recent analysis, Palomares-Ruiz and
Petcov\cite{petcov} have discussed the resolution of the issue of
hierarchy of neutrino masses in magnetised iron calorimeters by looking
at the ratio (and difference) of events with $\mu^+$ and $\mu^-$ due to
atmospheric muon neutrino and anti-neutrino interactions.

In principle an iron calorimeter ({\sc ical}) detector with charge
identification capability as has been proposed by the {\sc monolith}
\cite{monolith}, {\sc minos} \cite{minos} and {\sc ical/ino} \cite{ino}
collaborations, with a mass of at least 30--50 ktons, can resolve this
issue; the question at hand is one of feasibility and exposure times.

In this paper we focus on a new {\em difference asymmetry} that
is especially suited to probe the mass hierarchy. This asymmetry is
constructed from observables involving the detection of muon neutrinos
(and anti-neutrinos) through charged current interactions with muons in
the final state. In an earlier work \cite{petcov} it was shown that the
direct and inverted hierarchies could be discriminated by studying the
{\em magnitude} of certain asymmetries; in contrast, the new asymmetry
probes the hierarchy issue via its {\em sign}, which is surely a more
robust signature. Provided $\sin^2 2\theta_{13} > 0.04$, an {\sc ical}
detector is sensitive to the mass hierarchy. A result with a significance
greater than 90\% CL requires large exposures of more than 500 kton-years
as well as good energy and angular resolution of the detected muons
(better than 15\%), especially for small $\theta_{13}$. Hence obtaining
definitive results with such a detector is difficult, unless $\theta_{13}$
turns out to be large.

Furthermore, such an asymmetry is not as sensitive to the magnitude of
$\theta_{13}$; in the case of a positive result it may only place a lower
bound on this parameter. The issue of $\theta_{13}$ therefore may not be
resolved in such detectors using atmospheric neutrinos alone.

In section 2 we present some details of the oscillation probabilities and
their dependence on masses and mixing parameters after summarising the
status of their determination within the three neutrino frame-work. Some
features of these probabilities within some reasonable approximation
such as single-scale dominance are presented in a semi-analytic
treatment. The new asymmetry and its sensitivity to the hierarchy
are discussed here as also the crucial question of binning the data to
maximise this asymmetry. In section 3 we analyse the results numerically
for the event rates both together and separately for different charged
lepton final states using the {\sc nuance} code \cite{nuance} for atmospheric
neutrinos distributed according to the 3-D Honda flux \cite{honda}. We
discuss the sensitivity of the {\sc ical} detector to the hierarchy of
neutrino masses. This obviously depends on the mixing parameters. We
highlight the sensitivity to the two unknown parameters, the (13)
mixing angle and the sign of $\delta_{32}$. In section 4 we generate
events using the {\sc nuance} neutrino generator and study the impact of
statistical errors and fluctuations on the rates and asymmetries.
The effect of detector resolution and cuts on the observables is also
discussed. In section 5 we present our conclusions.

\section{Three flavour oscillations}

An understanding of the gamut of data from solar, atmospheric and reactor
neutrinos requires a mixing scheme with at least three neutrinos in
which the neutrino flavour states $|\nu_\alpha \rangle$, $(\alpha = e,
\mu, \tau)$ are linear superpositions of the neutrino mass eigenstates
$|\nu_i\rangle$, $(i=1,2,3)$, with masses $m_i$ :
\begin{equation}
\vert\nu_\alpha \rangle = \sum_i U_{\alpha i} \vert\nu_i\rangle~.
\end{equation}
Here $U$ is the $3 \times 3$ unitary matrix which may be parametrised 
as \cite{mixing} (ignoring Majorana phases):
\begin{eqnarray}
U = \left(
          \begin{array}{ccc}
          c_{12}c_{13} & s_{12}c_{13} & s_{13}e^{-i\delta}  \\
 -c_{23}s_{12} - s_{23}s_{13}c_{12}e^{i\delta} & c_{23}c_{12} -
s_{23}s_{13}s_{12}e^{i\delta}&  s_{23}c_{13}\\
  s_{23}s_{12} - c_{23}s_{13}c_{12}e^{i\delta}& -s_{23}c_{12} -
c_{23}s_{13}s_{12}e^{i\delta} & c_{23}c_{13} \end{array} \right).
\label{mns}
\end{eqnarray}
Here $c_{12}=\cos\theta_{12}$, $s_{12}=\sin\theta_{12}$ etc., and $\delta$
denotes the CP violating (Dirac) phase. By definition, the 3 $\times$
3 neutrino mass matrix $M_\nu$ is diagonalised in the charged-lepton
mass basis by $U$:
\begin{equation} 
U^\dagger M_\nu U = {\rm{diag}}(m_1,m_2,m_3). 
\label{dia}
\end{equation} 
The parameters involved are the mixing angles $\theta_{ij}$ and the
mass-squared differences $\delta_{ij} \equiv m_i^2 - m_j^2$. A combined
analysis \cite{analysis} of the data from all the experiments gives
the following best fits to the oscillation parameters:
\begin{itemize}

\item A combination of solar neutrino experiments and the KamLAND reactor 
experiment gives
\begin{eqnarray}
5.4\times 10^{-5}~\rm{eV^2} & < &  \delta_{21} < 9.4 \times 
10^{-5}~\rm{eV^2} \nonumber~, \\
0.30 & < & \tan^2 \theta_{12} < 0.64~,
\end{eqnarray}
with the best-fit values given by $\delta_{21} = 6.9 \times
10^{-5}~\rm{eV^2}$ and $\tan^2 \theta_{12} = 0.40 $. Note in particular
that $m_2^2 > m_1^2$.

\item A combination of atmospheric neutrino experiments and the K2K
experiment gives
\begin{eqnarray}
1.4 \times 10^{-3}~\rm{eV^2} & < & |\delta_{32}| < 5.1 \times 
10^{-3}~\rm{eV^2}~, \nonumber \\
\sin^2 2\theta_{23} & > & 0.86~,
\end{eqnarray}
with the best-fit values given by $|\delta_{32}| = 2.0 \times
10^{-3}~\rm{eV^2}$ and $\sin^2 2\theta_{23} = 1 $. Note that there is
as yet no determination of the sign of this mass-squared difference.

\item The {\sc chooz} bound on the effective mixing angle using the
above parametrisation is,
\begin{equation}
\sin^2 2\theta_{13} < 0.249 ~ \qquad \hbox{(99.73\% C.L.)}
\end{equation}
so that $\theta_{13} < 14.9^\circ$.
\end{itemize}

Inclusion of matter effects on the neutrino oscillation probabilities has a
complicated dependence on these parameters; furthermore, the
oscillation probability can only be computed numerically using a reference
model for the variation of earth's density. Before we present the exact
numerical results, we first discuss a simplified analytical model in
vacuum and matter.

\subsection{Probabilities in Vacuum}

\begin{figure}[htp]
\vskip 14truecm
{\includegraphics{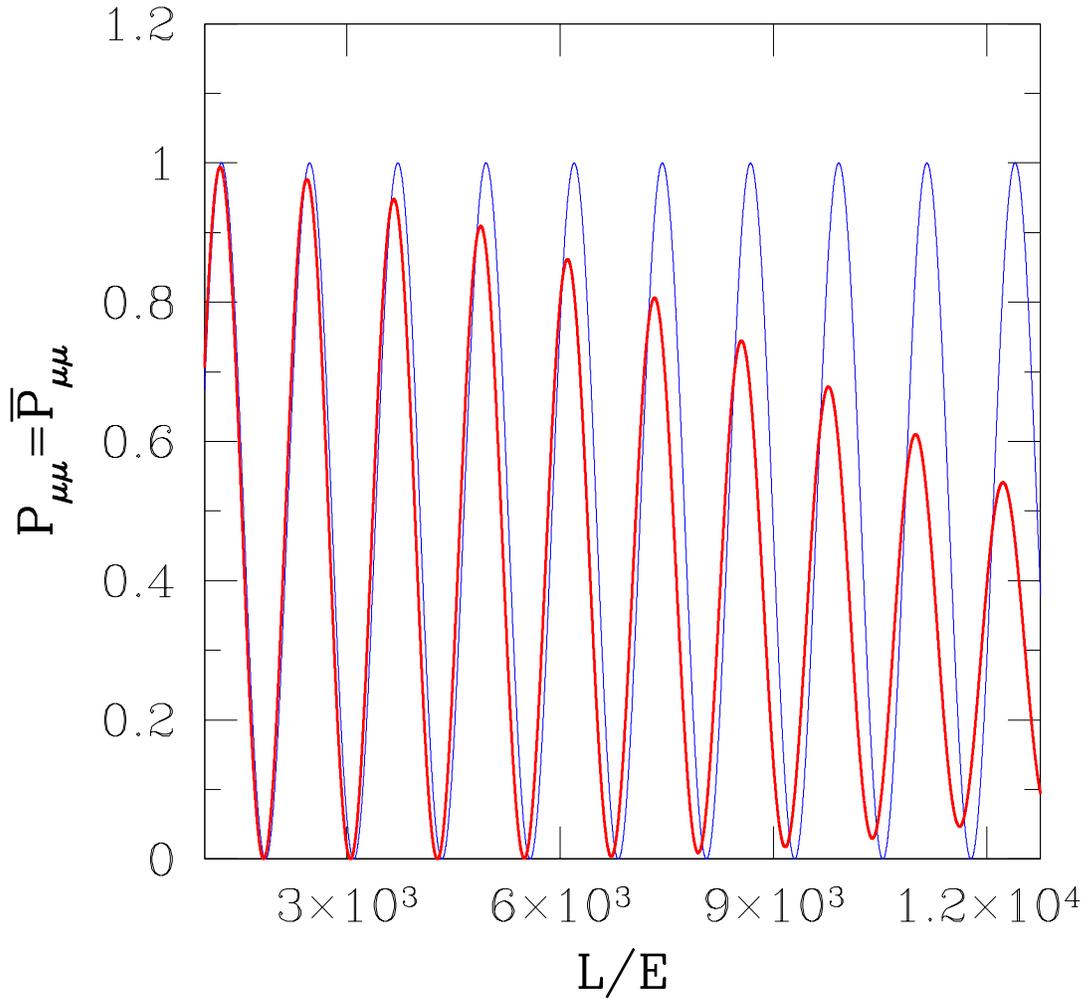}}
\caption{Survival probability of muon neutrinos in vacuum with two and 
three neutrino mixing scenarios. We have used $|\delta_{32}| = 2 \times 
10^{-3},~\theta_{23}=\pi/4$ in both. For the three generation probability we 
have used in addition $\theta_{13}=9^{o}$ and the best fit values of 
other parameters as given in the text.}
\label{vac} 
\end{figure}

In order to focus on the issues involved and their disentanglement we
first discuss the vacuum oscillations (actually the muon neutrino survival)
probabilities. The survival probability of a muon neutrino of energy $E$
(in GeV) after traversing a distance $L$ (in km) is given by
\begin{eqnarray}
P_{\mu\mu} & = &  1 - [2U_{\mu 1}U_{\mu 2}]^2 
\sin^2\left(\frac{1.27\delta_{21}L}{E}\right) - [2U_{\mu 2}U_{\mu 3}]^2
\sin^2\left(\frac{1.27\delta_{32}L}{E} \right) \nonumber \\
& & \qquad - [2U_{\mu 1}U_{\mu 3}]^2
\sin^2\left(\frac{1.27\delta_{31}L}{E} \right)~,
\label{p3nu}
\end{eqnarray} 
where we have neglected the CP violating phase. Note that in vacuum the
probabilities are the same for both neutrinos and anti-neutrinos. In
Fig.~\ref{vac} we plot the survival probability of $\nu_{\mu}$ as a
function of $L/E$ for $\delta_{32} = 2 \times 10^{-3}$ eV$^2$ and
$\theta_{13} = 9^\circ$, with other parameters fixed to their best-fit
values. For comparison we have also shown the two generation
oscillation probability,
\begin{equation}
P^{(2)}_{\mu\mu} = 1 - \sin^2 2\theta_{23}
\sin^2 \left(\frac{1.27\delta_{32}L}{E} \right)~,
\label{p2nu}
\end{equation} 
which has been widely used to fit $|\delta_{32}|$ and $\sin\theta_{23}$
from data.

Notice that the three generation formula reduces to the two generation
one only after setting $\theta_{13}=0$ and $\delta_{21}=0$. Even though
there are two oscillating terms (frequencies) controlling the survival
probability, the value of the probability at small $L/E$ is dominated
by the last two terms in eq.~\ref{p3nu}. In turn, this implies that the
first observable probability minimum is roughly independent of both
$\delta_{21}$ and $\theta_{13}$, even for $\theta_{13}$ as large as
the {\sc chooz} limit. At larger values of $L/E$, when $\delta_{21} L/E$
becomes large, the subdominant frequency controlled by $\delta_{21}$
has substantial effect.

This can be seen from Fig.~\ref{vac} where both the two-flavour and
three-flavour probabilities are plotted as a function of $L/E$. In fact,
for very large $L/E$, and for this typical set of oscillation parameters,
the vacuum probability $P_{\mu\mu}$ averages to $\approx 3/8$ and not
to $1/2$ as is the case for two generation mixing.

Notice that the probabilities in vacuum are the same for both
neutrinos and anti-neutrinos and depend only on the absolute value of
the mass-squared differences. However, the inclusion of three flavours
automatically turns on matter effects which alter these vacuum results;
we now examine the modification of these expressions due to matter
effects.

\subsection{Probabilities in matter}

In order to simplify the analysis, we first consider the propagation of
neutrinos and anti-neutrinos produced in the atmosphere in a slab of
constant density $\rho$ (in gms/cc)\cite{matprob}. In this case, we can
retain the same expressions as in vacuum for the probabilities, see
eq.~\ref{p3nu}, but replace the mass-squared differences and mixing
angles by the corresponding matter-modified effective values obtained by
diagonalising the matter dependent matrix (Hamiltonian):
\begin{eqnarray}
U \left(
          \begin{array}{ccc}
          0 & 0 & 0  \\
          0 & \delta_{21} & 0 \\
          0 & 0 & \delta_{31} \end{array} \right)U^{\dagger} 
 +\left(
          \begin{array}{ccc}
          A & 0 & 0  \\
          0 & 0 & 0  \\
          0 & 0 & 0 \end{array} \right),
\label{ham}
\end{eqnarray}
where
\begin{equation}
A = 2 \sqrt{2} G_F n_e E = 7.63 \times 10^{-5}~{\rm eV}^2~\rho({\rm gm/cc})~
E({\rm GeV})~ \hbox{eV}^2.
\label{densm}
\end{equation}
Here $G_F$ and $n_e$ are the Fermi constant and electron number density
in matter and $\rho$ is the matter density. Further simplification arises
because $\delta_{21} \ll \delta_{31}$ and we can treat the propagation
in matter as a one mass-scale problem involving only $\delta_{32}\approx
\delta_{31}$. The matter dependent mixing angle $\theta_{12,m}$ may be
approximately written as
\begin{equation}
\sin 2\theta_{12,m} \approx \frac{\sin 2\theta_{12}}{\sqrt{(\cos 
2\theta_{12} - (A/\delta_{21}) \cos^2\theta_{13})^2 + \sin^2 
2\theta_{12}}}~. 
\end{equation}
The condition for ``resonance" in the $(12)$ channel is given by
$\delta_{21} \cos 2\theta_{12} = A \cos ^2\theta_{13}$ which roughly
translates to the condition, $\rho E \approx 0.5$. We note in passing
that the resonance in the (12) sector for atmospheric neutrinos has a very
different character from its solar counterpart. For example, for energies
above 1 GeV, the resonance occurs at extremely small densities, even
in the atmosphere itself for some energies ! Furthermore, the resonance
width which is determined by $\sin^2 \theta_{12}$ is rather broad, with
the tail extending below $A = 0$ so that the effect of matter in this
sector on both neutrinos and anti-neutrinos will be roughly similar.

This also implies that $\sin 2\theta_{12,m}$ is rather small for neutrinos
or anti-neutrinos of few GeV energy even in low density regions such as
the earth's mantle---in fact at high densities and energies, we have
\begin{equation}
\sin 2\theta_{12,m} \approx \frac{\sin 2\theta_{12}}{\rho E}~,
\end{equation}
so that one can neglect the term depending on $\delta_{21}$ for
multi-GeV atmospheric neutrino propagation in earth matter. Indeed
the same argument also holds for anti-neutrinos since the equation
above\footnote{This term is often neglected citing the reason that
$\delta_{21}/\delta_{31} \ll 1$. Actually, $\delta_{21}L/E$ can be
of the order of unity for atmospheric neutrinos; it is the smallness
of the {\em coefficient} of this term that allows it to be dropped.}
holds for them as well.

The effect of matter on the angle $\theta_{13}$ is non-trivial 
and is given by
\begin{equation}
\sin 2\theta_{13,m} =\frac{\sin 2\theta_{13}}{\sqrt{( \cos 
2\theta_{13} - (A/\delta_{32}))^2 + (\sin 2\theta_{13})^2}}~.
\end{equation} 
The resonance condition in the $(13)$ channel is then given by
$\delta_{32} \cos 2\theta_{13} = A $ with a width $\delta A = \delta_{32}
\sin 2\theta_{13}$. Since $\sin^2 2\theta_{13}$ is small, the resonance is
rather sharp. This condition is satisfied for range of values $13 < \rho
E < 40 $ for $1 < \delta_{31}/10^{-3} < 3$ eV$^2$. The average mantle
density is about 4--5 gm/cc and the resonance occurs for energies greater
than 4 GeV. In the core, where the density exceeds 9 gm/cc the resonance
condition is satisfied with even smaller energy. For anti-neutrinos
the matter effects come with opposite sign, $A \rightarrow -A$, so that
resonance occurs with the inverted mass hierarchy $\delta_{31} < 0$.

The main observable effect due to propagation in matter comes from the 
matter dependent angle $\theta_{13,m}$ since for all practical purposes
we can set 
\begin{equation} 
\sin 2\theta_{23,m} \approx \sin 2\theta_{23}~.
\end{equation}
The approximate survival probability\cite{matprob} for $\nu_{\mu}$ can be 
written as 
\begin{eqnarray}
P^m_{\mu\mu}(A,\delta_{32}) & \approx & 1 - \sin^2 2\theta_{23} 
[\sin^2\theta_{13,m} 
\sin^2\Delta_{21}^m+\cos^2\theta_{13,m}\sin^2\Delta_{32}^m] \nonumber \\  
& & \qquad -\sin^2 2\theta_{13,m} 
\sin^4\theta_{23}\sin^2\Delta_{31}^m~,
\label{p3num}
\end{eqnarray} 
where
\begin{eqnarray}
\Delta_{21}^m & = & \frac{1.27 \delta_{32} L}{E}\frac{1}{2}\,
\left[\frac{\sin 2\theta_{13}}{\sin 2\theta_{13,m}} - 1 -
\frac{A}{\delta_{32}} \right]~, \nonumber \\
\Delta_{32}^m & = & \frac{1.27 \delta_{32} L}{E}\frac{1}{2}\,
\left[\frac{\sin 2\theta_{13}}{\sin 2\theta_{13,m}} + 1 +
\frac{A}{\delta_{32}} \right]~, \nonumber \\
\Delta_{31}^m & = & \frac{1.27 \delta_{32} L}{E}\,
\left[\frac{\sin 2\theta_{13}}{\sin 2\theta_{13,m}} \right]~. 
\end{eqnarray}
Note that for sufficiently large values of $A/\delta_{32}$, all the
three scales are of the same order of magnitude including 
$\Delta_{21}^m$
which cannot therefore be neglected.

If we further take $\sin^2 \theta_{13}$ to be  small, which it is, and 
restrict the analysis to values of $A$ away from resonance, then we can
expand the ratio
\begin{equation}
\by{\delta \sin 2\theta_{13}} {\sin 2\theta_{13,m}} \approx (\delta - A) 
+ \by{2 A \delta} {\delta - A}\,  \sin^2\theta_{13}~,
\end{equation} 
where we have used the notation $\delta \equiv \delta_{32}$.
As a result the survival probabilities for $\nu_{\mu}$ and 
$\bar{\nu}_{\mu}$ can be simplified further and written, to order
$\sin^2\theta_{13}$, as,
\begin{eqnarray} \nonumber
P^m_{\mu\mu}(A,\delta) & \approx & P^{(2)}_{\mu\mu} -
\sin^2\theta_{13} \left[ \by{A}{\delta-A} T_1 + 
\left(\by{\delta}{\delta-A}\right)^2
\left(T_2 \sin^2[(\delta -A) x] + T_3\right) \right]~, \\
\overline{P}^m_{\mu\mu}(A,\delta) & \approx & P^{(2)}_{\mu\mu} -
\sin^2\theta_{13} \left[ -\by{A}{\delta + A} T_1 + 
\left(\by{\delta}{\delta+A}\right)^2
\left(T_2 \sin^2[(\delta +A) x] + T_3\right) \right]~,
\label{pmm}
\end{eqnarray} 
where $x \equiv 1.27 L/E$. Here $P^{(2)}_{\mu\mu} = 1 -
\sin^2 2\theta_{23} \sin^2 (x \delta)$ is the
2-generation muon neutrino survival probability that is independent of
matter effects as well as $\sin^2\theta_{13}$. The matter
dependence appears in the remaining terms, where
\begin{eqnarray} \nonumber
T_1 & = & x \delta \sin^2 2\theta_{23} \sin 2 \delta x~, \\ \nonumber
T_2 & = & 4 \sin^4 \theta_{23} ~, \\ \nonumber
T_3 & = & \sin^2 2\theta_{23} \left(\sin^2 A x - \sin^2 \delta x \right)~. 
\end{eqnarray} 
Notice that the survival probability for the anti-neutrinos,
$\overline{P}^m_{\mu\mu}$, is given by the replacement $A \to -A$ in
$P^m_{\mu\mu}$. Both probabilities are invariant under $(A, \delta)
\leftrightarrow (-A, -\delta)$.

Several important points emerge from this analysis:
\begin{itemize}

\item As was noticed in the analysis of the vacuum probabilities
with three flavours, there are two mass scales and their associated
frequencies.  Even though one of the scales is small in magnitude,
$\delta_{21}\ll \delta_{31}$, it does make an impact on the survival
probability due to the large $L/E$ available; this happens because a
large range of $L/E$ is available for atmospheric neutrinos. This makes
the 3-flavour results differ strongly from the 2-flavour ones in the
large $L/E$ region, in the vacuum case.

\item The matter effect however effectively removes one of these
frequencies controlled by $\delta_{21}$ such that the oscillations
are controlled by the frequency determined by $\delta_{32}$ and to a
lesser extent by the matter term $A$. In other words, the deviation
of the 3-flavour probability from the 2-flavour one at large $L/E$
disappears when matter effects are turned on. The residual matter effect
in atmospheric neutrinos is thus a sub-leading effect.

\item Thus the matter dependent $P^m_{\mu\mu}$ and
$\overline{P}^{\,m}_{\mu\mu}$ resemble the two-generation probability
curve shown in Fig.~\ref{vac} rather than the three generation one.
In fact the first dip is almost completely unaltered by the matter
effects, and the average behaviour for large $L/E$ is similar to the
two-generation case.  This in fact justifies the attempts to get bounds
on $|\delta_{32}|$ and $\sin 2\theta_{23}$ using fits to the Super-K
data within the two-flavour oscillation formalism.

\item Therefore the sensitivity to $\theta_{13}$ and the sign of
$\delta_{32}$ are hidden in the sub-leading effects which show up in
the oscillation frequency, the width and the amplitude after the first
dip, and are maximal for intermediate values of $L/E$. The matter effect
is more pronounced for (anti-)neutrinos in the case of the (inverted)
direct hierarchy, since the amplitude of the matter-dependent term is
enhanced (decreased) when $\delta_{32}$ and $A$ have the same
(opposite) sign, as can be seen from eq.~\ref{pmm}.

\end{itemize}

All of the above statements apply appropriately (by interchanging the
matter effects for neutrinos and anti-neutrinos) for both the direct
and inverted hierarchies. This {\it exchange symmetry} will however be
broken once the cross-sections are taken into account since the
neutrino cross-sections are on the average a factor of two more than
the anti-neutrino cross-sections. Since matter effects are enhanced for
neutrinos in the case of the direct hierarchy, such a hierarchy will be
more favourable for detection due to the larger events sample for the
same exposure time.

\section{Event rates and asymmetries}

The arguments of the previous section clearly indicate that the main
effect of matter on the survival probability is to reduce it to a single
frequency (which however is not a constant since it depends both on energy
and the density) oscillation, whereas in vacuum there are clearly two
well-separated oscillations.

The inclusion of matter effects through a 3-flavour analysis is
complicated by an additional energy dependence.  While the survival
probability in the 2-flavour case depends only on the ratio $x = L/E$,
that in 3-flavours depends explicitly on both $x$ and $E$, the latter
through the matter term. The experimental observable is the event rate
in the detector. The fully differential event rate for neutrinos of
flavour $\alpha$ to be detected is given by the general expression:
\begin{equation}
\frac{\d N^{\alpha}}{\d\ln E\d x} = 
K_y\sum_{\beta} P_{\beta\alpha}(E,x)~\Phi_\beta(E, x) ~ \sigma_{\alpha}(E)~,
\label{eventr}
\end{equation} 
where $x = L/E$ and $\sigma_{\alpha}$ is the total interaction
cross-section for the $\alpha$ type neutrino to interact with the
detector material. Here $P_{\beta\alpha}$ is the conversion probability
of a neutrino of flavour $\beta$ to a flavour $\alpha$.

The flux-dependent term $\Phi_\alpha (E, x)$ is obtained from the
doubly-differential neutrino (or anti-neutrino) flux of flavour $\alpha$,
$\d^2 \phi_{\alpha}(E,z)/\d \ln E \d z$, which is a function of the
energy $E$ and zenith (actually nadir) angle $z=\cos\theta$, by
multiplying with a suitable Jacobean factor.

The factor $K_y$ is the detector dependent factor measured in units of
kton-years. The detector is assumed to be mainly made up of magnetised
iron with active detector elements. In the {\sc monolith} and {\sc
ical/ino} proposals, the active detector elements are glass resistive
plate gas filled chambers (RPCs). In either of these proposals the
detector mass is almost entirely ($>$ 98\%) due to its iron content. We
will be interested here in {\em event ratios}; hence the factor $K_y$
and other actual detector details are not necessary for the analysis in
this section.  However, we assume that the detector is capable of
identifying the charge of the muons in the final state with good
precision.

The distance of propagation $L$ of the neutrino from the point of
production to the detector is obtained from
\begin{equation}
L = \sqrt{(R+L_0)^2-(R\sin\theta)^2} -  R\cos\theta~,
\label{Ldef}
\end{equation}
where $\theta=0$ corresponds to neutrinos reaching the detector vertically
downwards after a distance $L_0$ which is the average height above the
surface of the earth at which the atmospheric neutrinos are produced. We
take this to be about 15 kms, as is the convention. Here $R$ is the
radius of the earth.

The event rate in a given bin of $x = L/E$ is,
\begin{equation}
N^{\alpha}_{\rm bin}(x) = \int_{\rm bin}  \d x \int_{E_{\rm min}} \by{\d E}{E}
\frac{\d^2 N^{\alpha}}{\d\ln E\d x}~;
\label{eventb}
\end{equation}
henceforth we discuss only the case of muon-neutrinos, $\alpha= \mu$.
For the case of atmospheric neutrinos of interest here, both $P_{e\mu}$
and $P_{\mu\mu}$ contribute in the expression above. However, due to the
smallness of the (13) mixing angle, $\theta_{13}$, the contribution of
$P_{e\mu}$ is very small and is maximum near large $L/E$ where it is
about 10--20\% (5\%) for neutrinos (anti-neutrinos).

The event rate is expressed as a function of $x$, averaged over a bin
width that will be appropriately chosen to maximise the sensitivity to
the sign of $\delta_{32}$. The expression given in eq.~\ref{eventb}
is the best case scenario since the integration is over the neutrino
or anti-neutrino energy. We discuss the effect of including the detector
resolution functions in the next section.

In order to proceed further we also distinguish between the up-coming
and down-going neutrinos (and anti-neutrinos). It is obvious that the
down-going particles travel a much smaller path-length of $15 < L < 400$
km in matter. Hence effects of oscillation are expected to be negligible
for these neutrinos\footnote{This is not true for very horizontal
neutrinos; however, if the detector geometry consists of horizontally
stacked iron-plates, such very horizontal events cannot be observed.}.

\begin{figure}[htp]
\vskip 11 truecm
\includegraphics{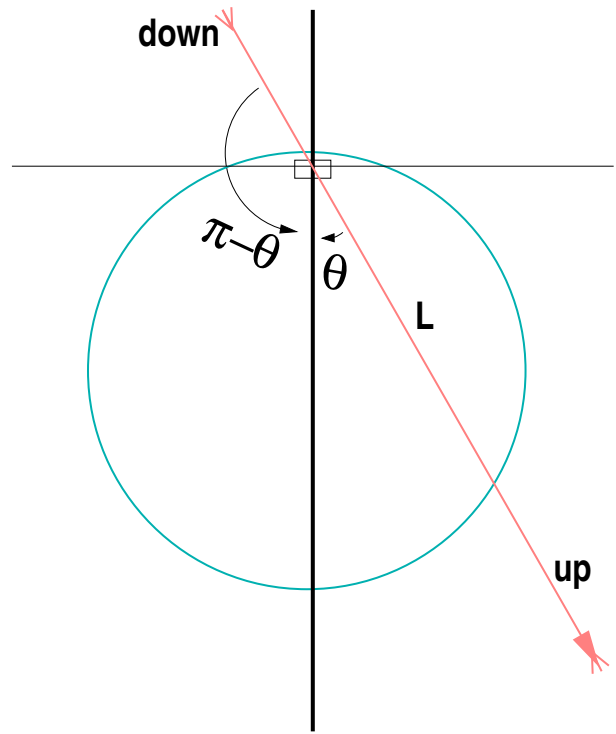}
\caption{A schematic of the relation between up-coming and down-going
neutrinos with $\theta \leftrightarrow (\pi - \theta)$.}
\label{updown} 
\end{figure}

A useful measure of oscillations is the ratio of up-coming to down-going
neutrinos with nadir/zenith angles interchanged. This is clear from
Fig.~\ref{updown}. The fluxes of atmospheric neutrinos from directions
$\theta$ and $(\pi - \theta)$ are expected to be similar in the absence
of oscillations, especially for larger energies, $E >$ few GeV. Since
the path-length traversed, $L$, is related to $\theta$ as
$$
L = f(\vert \cos\theta \vert) - R \cos\theta~,
$$
(see eq.~\ref{Ldef}), the replacement $\theta \leftrightarrow (\pi -
\theta)$ effectively changes the sign of the second term in the equation
above, thus taking, for instance, a down-going neutrino to an up-coming
one. The ratio of events in the up-down directions for a given $x =
L/E$, therefore, reflects the asymmetry of the up-down fluxes, due to
oscillations. We define \cite{picchi}
$$
{\cal{R}} = \by{U}{D}(x) =
\by{\hbox{No. of events from up-coming muon neutrinos} (x)}
{\hbox{No. of events from down-going muon neutrinos} (\tilde{x})}~,
$$
where $\tilde{x} = x (\theta \leftrightarrow (\pi - \theta))$ and the
number of up-coming (U) or down-going (D) events is calculated using
eq.~\ref{eventb}. Similarly $\overline{U}/\overline{D}$ defines the
corresponding up-down ratio for anti-neutrinos. Since the effect of
oscillations on the denominator is small, the ratio ${\cal{R}}$ is
effectively the ratio of oscillated to unoscillated events with the same
$L/E$. 

Thus, while reflecting the effect of oscillations, this ratio
also minimises errors due to the uncertainties in the overall flux
normalisation (which can be as large as 30\%). Such a ratio, and
its sensitivity to the parameters of neutrino oscillations, has been
well-studied before; in particular, detailed studies have been carried out
by the {\sc monolith} collaboration \cite{monolith}. We shall not repeat
the results of such a study here; however, we will use the information
on the absolute value of $\delta_{32}$ that such a study will yield in
order to analyse matter effects and its implication for the neutrino
mass hierarchy.

We now proceed with the numerical details. All calculations presented
are done using the {\sc nuance} neutrino generator. The atmospheric neutrino
flux given by Honda et al. \cite{honda} is used. The code generates
oscillation probabilities based on the three neutrino oscillation
framework of Barger et al. \cite{barger}. The earth matter effects are
calculated using the Preliminary Reference Earth Model (PREM)
\cite{prem} for the variation in matter density.

Apart from $L$ and $E$, the probabilities depend on the parameters
$\delta_{21}$, $\delta_{32}$, $\theta_{12}$, $\theta_{23}$, and
$\theta_{13}$ (the CP phase has been set to zero). Since the focus of
this paper is the issue of matter effects and mass hierarchy, we fix
the other parameters at their best-fit values\footnote{These values
have altered somewhat with the latest results from Super-K and KamLAND;
however, they should not affect our conclusions or even our numerical
calculations significantly.}: $$ \delta_{21} = 7.0 \times 10^{-5}
\rm{eV^2}~;~ \tan^2 \theta_{12} = 0.39~;~ \sin^2 2\theta_{23}=1.0~,
$$ and vary only the parameters of interest, viz., $\delta_{32}$ (both
in magnitude and in sign) and $\theta_{13}$. We restrict ourselves to
the latest Super-K limits on the magnitude of $\delta_{32}$: $ 1 \le
\vert \delta_{32}/10^{-3} \vert \le 3$ eV${}^2$ and the {\sc chooz}
bound $\sin^2 2 \theta_{13} < 0.249 (\theta_{13} < 14.9^\circ$).

The {\sc nuance} neutrino generator calculates the probabilities, fluxes
and cross-sections suitably in bins of $E$ and $\cos\theta$ for use in
generating events. Event rates, as given in eq.~\ref{eventr}, were first
computed using these tables, with suitable interpolation and integration
over the variables with a cut on the energy, $E_{\rm min} = 4$ GeV. This
cut was optimised to maximise the matter-dependent effects, as we shall
see below.

\begin{figure} [htp]
\vskip 14truecm
{\includegraphics{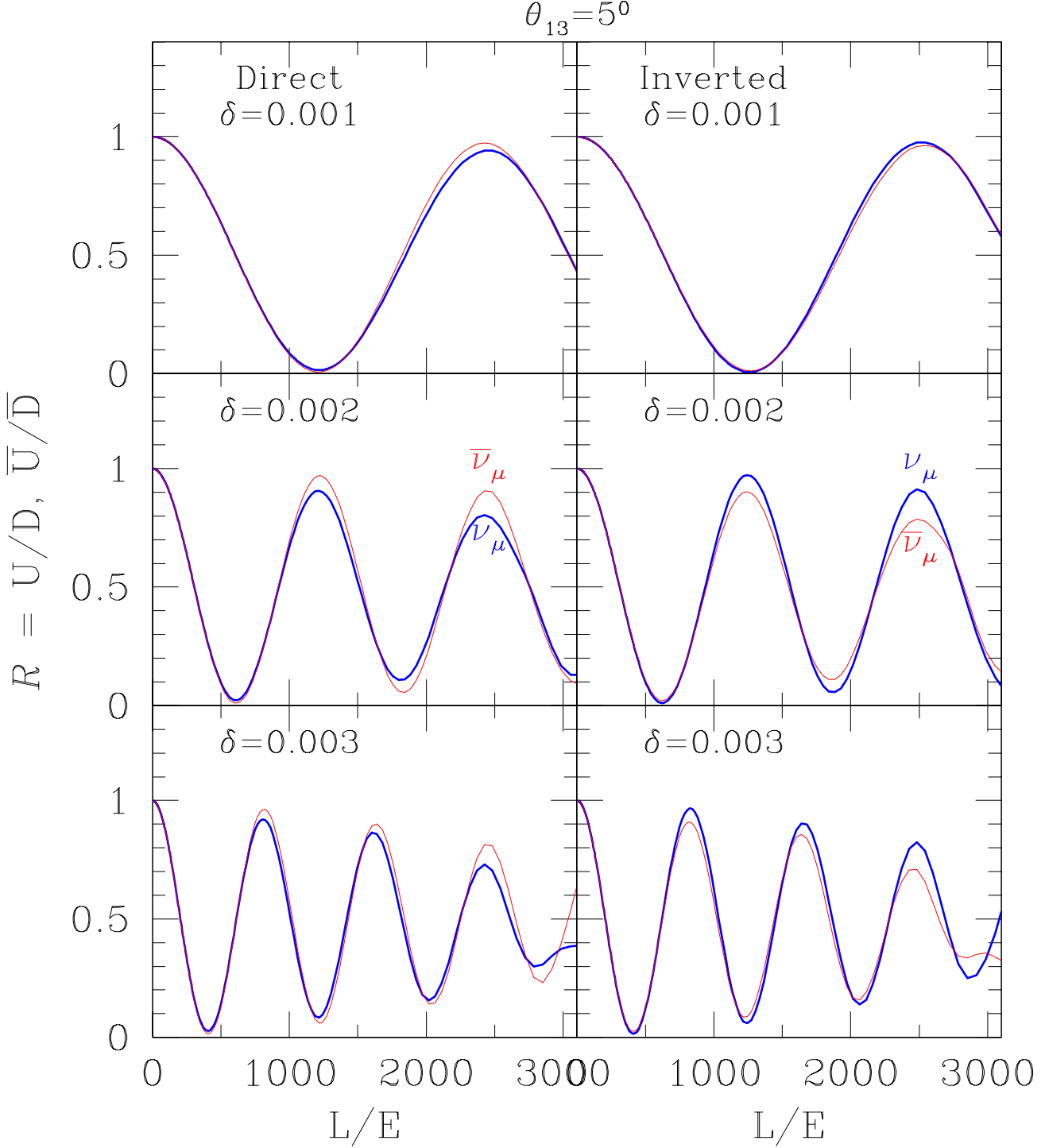}}
\caption{Up/down event ratios for $\nu_{\mu}$ shown as thick (blue) lines
and $\bar{\nu}_{\mu}$ as thin (red) lines as a function of $L/E$ for
$E>4$ GeV. The three horizontal panels correspond to $\delta \equiv
|\delta_{32}| = 1, 2, 3 \times 10^{-3}$ eV${}^2$. The vertical
panels correspond to the direct ($\delta_{32} > 0$) and inverted
($\delta_{32} < 0$) hierarchy for $\theta_{13}=5^{\circ}$. All other
parameters are kept at their best fit values as given in the text.}

\label{le5} 
\end{figure}

\begin{figure}[htp]
\vskip 14truecm
{\includegraphics{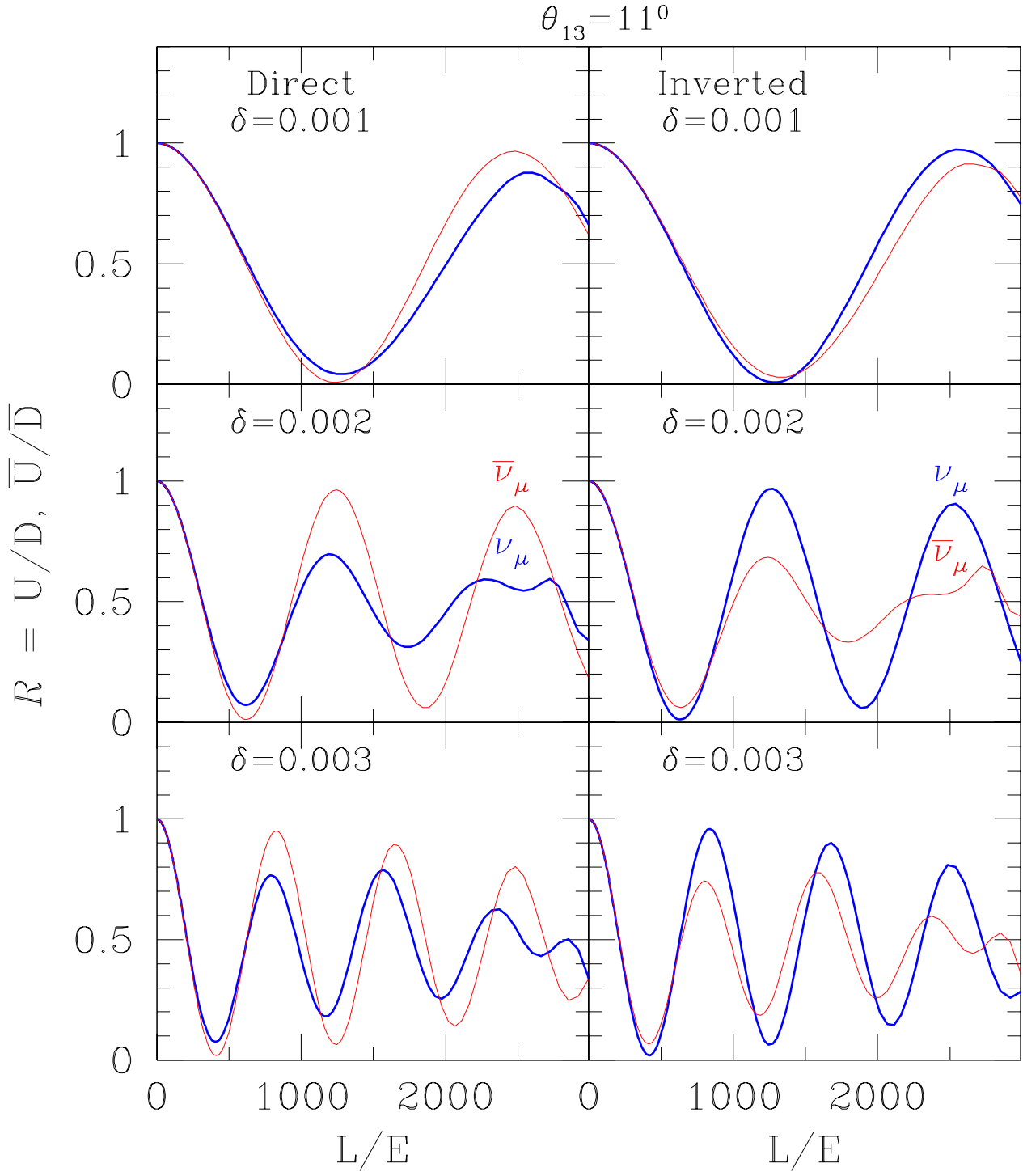}}
\caption{Same as in Fig.\ref{le5} with $\theta_{13}=11^{o}$.} 
\label{le11} 
\end{figure}

We show the variation of the up/down events ratio ${\cal{R}}$ as a
function of $L/E$ for both $\nu_\mu$ and $\overline{\nu}_\mu$ in
Figs.~\ref{le5} and \ref{le11} for direct ($m_3^2 > m_2^2$) and inverted
($m_3^2 < m_2^2$) hierarchies. Results are shown for $|\delta_{32}|=1, 2,
3 \times 10^{-3}$ eV${}^2$ and for 
$\theta_{13}=5^{\circ}$(Fig.~\ref{le5}) and $11^\circ$(Fig.~\ref{le11})
to show the sensitivity to these two parameters.

It is obvious from Figs.~\ref{le5} and \ref{le11} that there is very
little sensitivity to the sign of $\delta_{32}$ at small values of
$\theta_{13}$. For example, it is clear that atmospheric neutrino
experiments can say very little about the hierarchy problem below
$\theta_{13} = 6^\circ$ ($\sin^22\theta_{13} = 0.04$). However, the
sensitivity to the sign increases with the magnitude of $\theta_{13}$.

It is important to note that the position of the first minimum is
{\em independent} of both $\theta_{13}$ and the sign of $\delta_{32}$ but
depends only on the magnitude of the latter. Hence it can be reliably
used to determine the magnitude of $\delta_{32}$ as a precursor to
determining its sign with the same experimental set-up.

\begin{figure}[htp]
\vskip 14truecm
{\includegraphics{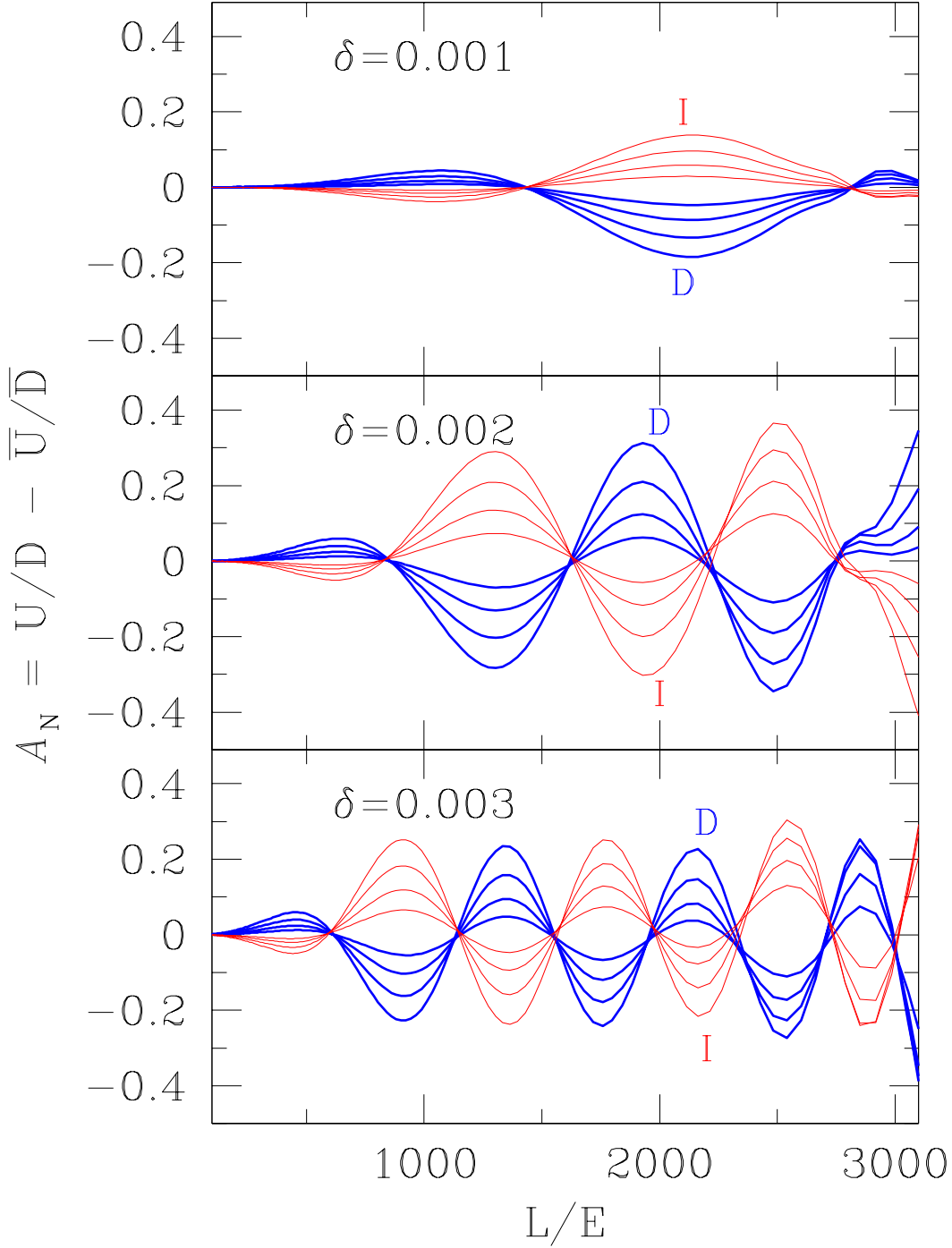}}
\caption{The difference asymmetry (difference of up/down ratios of
muon neutrinos and anti-neutrinos) as a function of $L/E$ for $E > 4$ GeV.
The three panels correspond to $\delta \equiv |\delta_{32}| = 1, 2,
3 \times 10^{-3}$ eV$^2$. The thick (blue) curves correspond to the
direct $\delta_{32} > 0$ and the thin (red) curves correspond to the
inverse $\delta_{32} < 0$ hierarchy. The innermost curve in each envelope
corresponds to $\theta_{13} = 5^{\circ}$ and the outermost corresponds to
$\theta_{13} = 11^\circ$ with $7^\circ$ and $9^\circ$ in between. The
best-fit values of other parameters are given in the text.}

\label{ledif}  
\end{figure}

It is clear that the neutrino and anti-neutrino up/down event ratios are
different from each other as well as different with direct and inverted
mass hierarchies (due to matter effects); the distinction (and hence
measurement possibilities) between the two hierarchies can be amplified
by defining the asymmetry,
\begin{equation}
{\cal{A}}_N(x) = \frac{U}{D}(x)-\frac{\overline{U}}{\overline{D}}(x)~.
\label{rateasym}
\end{equation}
The asymmetry, calculated numerically, and integrated over $E_{\rm
min} > 4$ GeV is plotted as a function of $L/E$ in Fig.~\ref{ledif}
for $\delta_{32} = 1, 2, 3 \times 10^{-3}$ eV${^2}$. The thick (blue) 
curves in the figure correspond to the direct mass hierarchy (labelled D)
and the thin (red) curves to the inverted mass hierarchy (labelled I).  
The
curves in each envelope correspond to $\theta_{13}=5,7,9,11$
degrees ($\sin^22\theta_{13}$ from 0.03--0.14) with the asymmetry
increasing symmetrically with $\theta_{13}$ about the ${\mathcal{A}}_N=0$
line for direct and inverse hierarchies. It is seen that the direct and
inverted asymmetries are exactly out of phase. 

The maximum divergence between the direct and inverted hierarchies
is smaller in the first envelope than in the second; these correspond
to the first dip and rise in the up/down events ratio (equivalently, in
$P_{\mu\mu}$). We will see in the next section that the statistically
significant region with maximum number of events corresponds to these
first few envelopes, i.e., to the range $500 < L/E < 1500$ for all the
three values of the $\delta_{32}$ considered here.

For neutrinos of energy about 5--7 GeV the interesting region thus
corresponds to distances 4000--10000 km with an average around 7000 km
\cite{magic} which is really the magic baseline where matter effects are
largest and the effect of the CP violating phase is negligible (though
we have set this phase to zero).

In order to understand why the direct and inverted asymmetries are out
of phase, it is convenient to appeal to the analytic expressions for the
probabilities discussed in the previous section. 

Since the events in a given $L/E$ bin are largely saturated by
the smallest allowed energy in that bin (the flux falls faster than
$1/E^2$), the effect of the cross-section factor is small and the up/down
events ratio is well-approximated by the flux-averaged probability
$P_{\alpha\beta}^{\rm av}$ defined below:
\begin{equation}
P^{\rm av}_{\alpha\beta}(\langle x\rangle) = \by{{\int_{\rm bin} \d x 
\int^{\infty}_{E_{\rm min}} \by{\d E}{E}~P_{\alpha\beta}(E,x)~
\Phi_\alpha(E, x)}} {{\int_{\rm bin}\d x \int^{\infty}_{E_{\rm min}}
\by{\d E}{E}~\Phi_\alpha(E, x)}}~.
\label{pavm}
\end{equation}
Considering only the dominant contribution from $P_{\mu\mu}$ to this
expression, we see that ${\cal{A}}_N$ corresponds to the approximate difference
asymmetry,
\begin{equation}
{\cal{A}}_N \approx {\mathcal{A}}(A,\delta_{32}) \equiv P^{\rm av}_{\mu\mu}-
\overline{P}^{\rm av}_{\mu\mu}~.
\label{asym}
\end{equation}
Here $\overline{P}$ denotes the survival probability for anti-neutrinos
which is obtained from the corresponding neutrino survival probability
$P$ by the replacement $A \rightarrow -A$ and the explicit dependences on
the energy and path length  $E$ and $L$ have been suppressed for clarity.

In the single mass-scale dominance approximation (with $\delta_{21}
\ll \delta_{31}, A$), the survival probability for neutrinos and
anti-neutrinos involves only the ratio $A/\delta_{32}$. As a result
we have,
\begin{eqnarray}
P^{\rm av}_{\mu\mu}(A,\delta_{32}) = 
P^{\rm av}_{\mu\mu}(-A,-\delta_{32})~;\nonumber \\ 
P^{\rm av}_{\mu\mu}(A,-\delta_{32})=  
P^{\rm av}_{\mu\mu}(-A,\delta_{32})~.
\end{eqnarray}
This is easy to see from the approximate expressions for the case of
constant-density matter in eq.~\ref{p3num}. It follows therefore, for
a given $E$ and $L$,
\begin{equation}
{\mathcal{A}}(A,\delta_{32}) = 
-{\mathcal{A}}(A,-\delta_{32})~,
\end{equation}
for constant matter density. Note that the amplitude is not a
constant but depends on the energy and distance travelled. In other
words, the asymmetry oscillates with $L/E$ while being exactly out of
phase for the direct and inverse hierarchies.  It appears (as can be
seen from Figs.~\ref{le5} and \ref{le11}) that this feature survives
in the events ratio computed using the exact formula without using the
approximation defined in eq.~\ref{asym} above.

Note also that the oscillation wavelength is a complicated function of
both the matter-dependent terms as well as the mass-squared difference,
$\delta_{32}$. However, the matter effect is not so significant in the
case of (anti-)neutrinos with (direct) inverted hierarchy, where the
oscillation extrema are essentially determined by the matter-independent
condition,
$$
\by{1.267 \vert \delta_{32} \vert L}{E} = n \by{\pi}{2}. 
$$

\begin{figure}[htp]
\vskip 14truecm
{\includegraphics{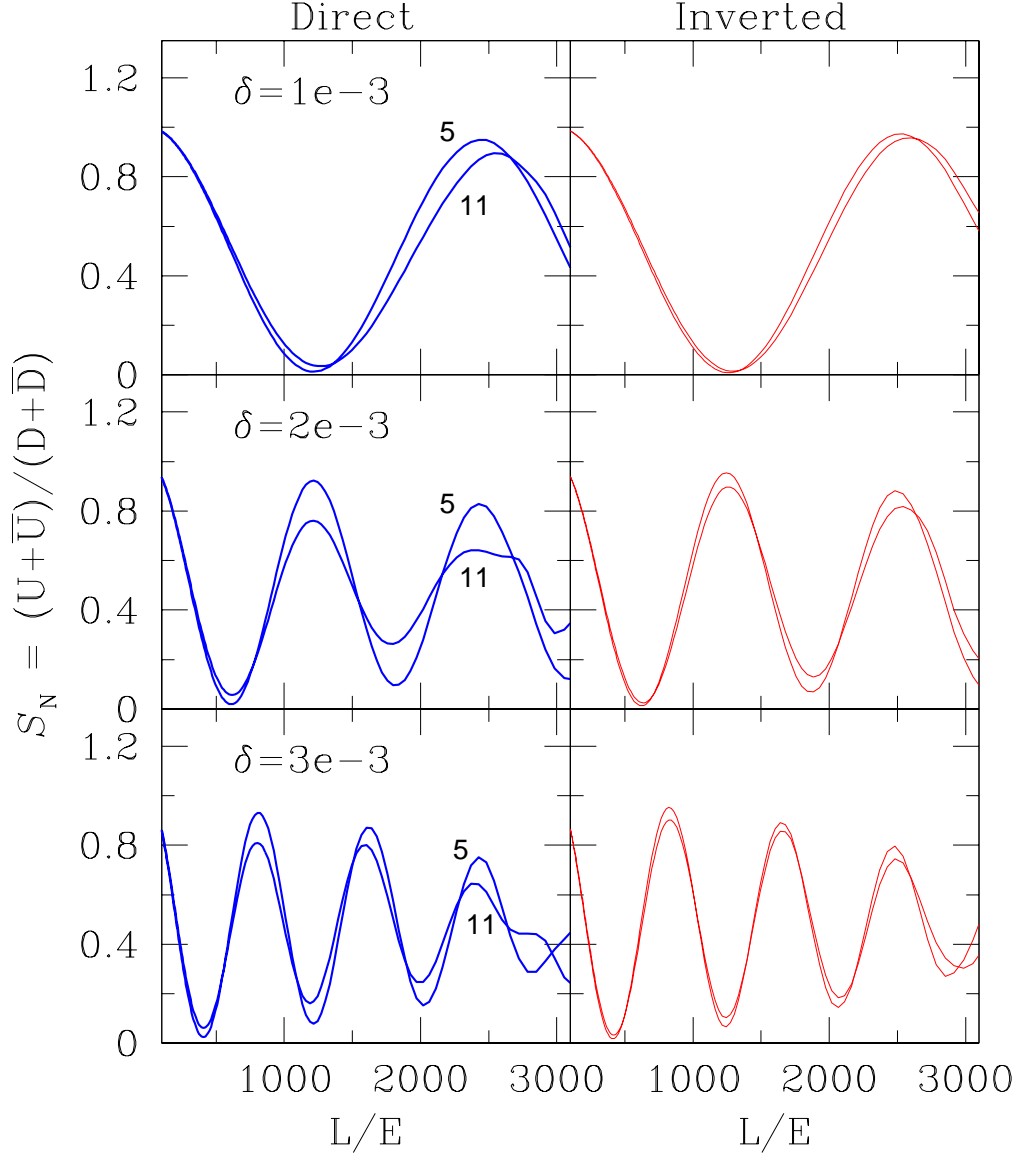}}
\caption{The total up/down ratio including both muon neutrinos and
anti-neutrinos as a function of $L/E$ for $E > 4$ GeV. The three panels
correspond to $\delta \equiv |\delta_{32}| = 1, 2, 3 \times 10^{-3}$
eV$^2$. The best-fit values of other parameters are given in the text.
The thick (blue) curves correspond to the direct ($\delta_{32} > 0$)
and the thin (red) curves correspond to the inverted ($\delta_{32} <
0$) hierarchy. The curves correspond to $\theta_{13}=5^{\circ}$ and
$11^\circ$ as indicated. Curves for in-between values of $\theta_{13}$
will lie in-between. Matter effects are more pronounced in the direct
hierarchy.}

\label{letot} 
\end{figure}

So far, we have discussed the asymmetry where separation of neutrino and
anti-neutrino events is a must. This can be arranged through observation
of charged-current channels with either $\mu^-$ or $\mu^+$ in the final
state, for instance, through the use of a magnetic field. However, in
low-counting experiments such as these, it is also useful to ask whether
any information can be gained from putting the neutrino and
anti-neutrino samples together, thus increasing statistics. We define
\begin{equation}
{\cal{S}}_N(x) = \frac{U + \overline{U}}{D + \overline{D}}(x)~.
\label{ratesum}
\end{equation}
Note that this does not need charge identification.
We show the dependence of ${{\cal{S}}_N}$ on the parameters $\delta_{32}$
(both magnitude and sign) and $\theta_{13}$ in Fig.~\ref{letot}. 
We see that the amplitude of
oscillation depends weakly on $\theta_{13}$ and that it dies
down faster for the direct hierarchy than for the inverted hierarchy. It
may therefore be possible to determine the mass hierarchy with the sum
of neutrino and anti-neutrino data, but for some-what larger values of
the (13) mixing angle, $\theta_{13} \ge 7^\circ$.

To sum up, the analysis of the up/down events ratio, their difference
and their sum, clearly indicates that 
\begin{enumerate}
\item the location of the first dip (for neutrinos and anti-neutrinos
with either hierarchy) depends only on the {\em magnitude} of $\delta_{32}$.

\item the $\nu_\mu - \overline{\nu}_\mu$ difference asymmetry is almost
exactly out of phase for direct and inverted hierarchies in $L/E$ bins
whose widths can be optimally determined by a {\em matter-independent}
condition involving $\vert \delta_{32}\vert$ and $L/E$.

\item the {\em amplitude} of the asymmetry defined above is clearly a
sensitive function of $\theta_{13}$.

\end{enumerate}

In the next section we examine the event rates and asymmetries in
a realistic magnetised iron calorimeter detector capable of charge
identification such as in the {\sc monolith}, {\sc minos} and {\sc
ical/ino} proposals.

\section{Event Rates with a neutrino generator}

So far we have discussed the theoretical up/down events ratios, a
difference asymmetry and its matter sensitivity. We now need to address
the question of statistics, i.e., estimate the detector dimension and the
time required for data accumulation. We generate events using the {\sc
nuance}
neutrino generator \cite{nuance} with the {\sc ical/ino} geometry which
consists of 140 layers of 6 cm thick iron plates of transverse section 16
m $\times$ 32 m. The RPCs are sandwiched between the layers. The fiducial
mass of the detector is a little under 30 ktons. Results shown correspond
to 480 kton-years data and indicate (1) the statistical significance of
the events sample (the error bars) and (2) the impact of fluctuations
(deviation of the central values from the predictions of the previous
section).

The $E$ and $L$ of those neutrinos that undergo charged current
interaction in the detector to produce muons in the final state were
directly used to generate the sums and asymmetries. The effect of finite
$L/E$ resolution is discussed in the next section. A detailed study
reconstructing the neutrino energy and direction from the final state
muons and hadrons is in progress.

As stated earlier, it was found that a cut of $E_{\rm min} = 4$ GeV
maximises the relative matter-dependent effect between neutrinos
and anti-neutrinos. While such effects are washed out beyond $E
\sim 10$ GeV, we integrated over all possible energies in a bin to
improve statistics. Data samples were generated for different values
of $\theta_{13}$ and $\delta_{32}$ (both signs), keeping the other
parameters fixed to the values listed in the previous section.

\subsection{The asymmetry}
To maximise the asymmetry, it is necessary to integrate the
events in a bin size that includes one half-period of the oscillation,
where the asymmetry is always positive or negative (see
Fig.~\ref{ledif}). Unfortunately, this is not easy since different path
lengths correspond to different matter densities and the wavelength is
density dependent. However, recall that the oscillation wave-length
was roughly constant and matter-independent for (anti-)neutrinos with
(direct) inverted hierarchy. We therefore identify the bins spanning
those $L/E$ corresponding to
\begin{equation}
\frac{1.267 \vert \delta_{32} \vert L}{E} = n \frac{\pi}{4}~; \quad n
\hbox{ odd}.
\label{bins}
\end{equation}
This has the advantage of being matter-independent. Once $\vert
\delta_{32} \vert$ is measured from the (matter/hierarchy-independent)
first minimum in the plot of the $U/D$ ratio as a function of $L/E$,
this can be used to generate the bin sizes for studying the
matter-dependent asymmetry.

\begin{figure}[htp]
\vskip 14truecm
{\includegraphics{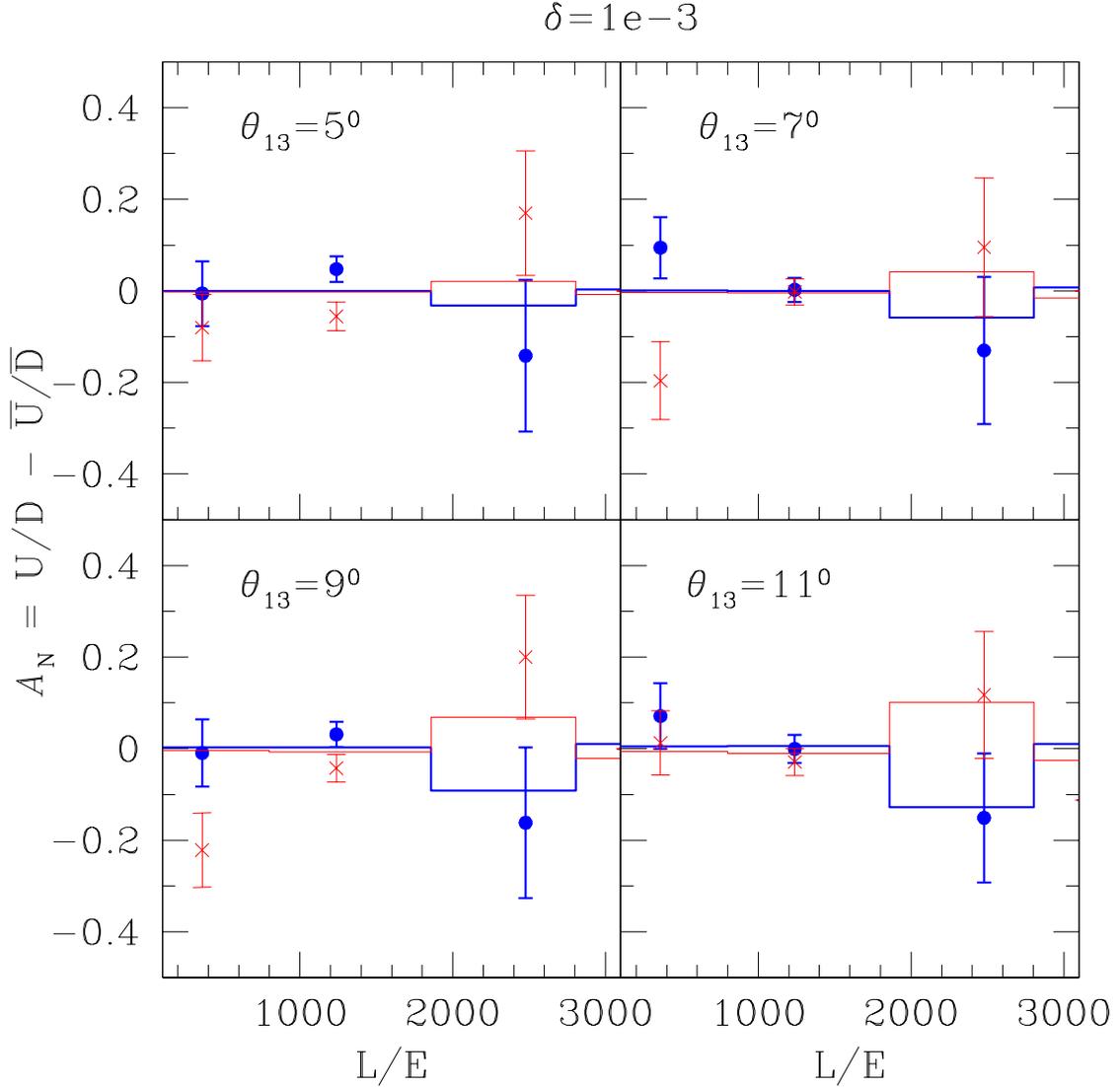}}
\caption{The up/down Monte Carlo events asymmetry ${\cal{A}}_N$ integrated
over $L/E$ bins shown as a function of $L/E$. The dots and crosses
correspond to the direct and inverted hierarchies respectively. Error bars
shown are obtained from the number of events in each bin. The histograms
correspond to the same asymmetry calculated from theory; see
Fig.~\ref{ledif}. Thick (blue) lines refer to the direct hierarchy
and thin (red) ones to the inverted one. The effect of fluctuations is
indicated by the shift in the central data values from the histogram.
Results are shown for $\theta_{13} = 5, 7, 9, 11$ degrees for $\vert
\delta_{32} \vert = 1 \times 10^{-3}$ eV${}^2$. For more details on the
bins, see the text.}

\label{le1} 
\end{figure}

\begin{figure}[htp]
\vskip 14truecm
{\includegraphics{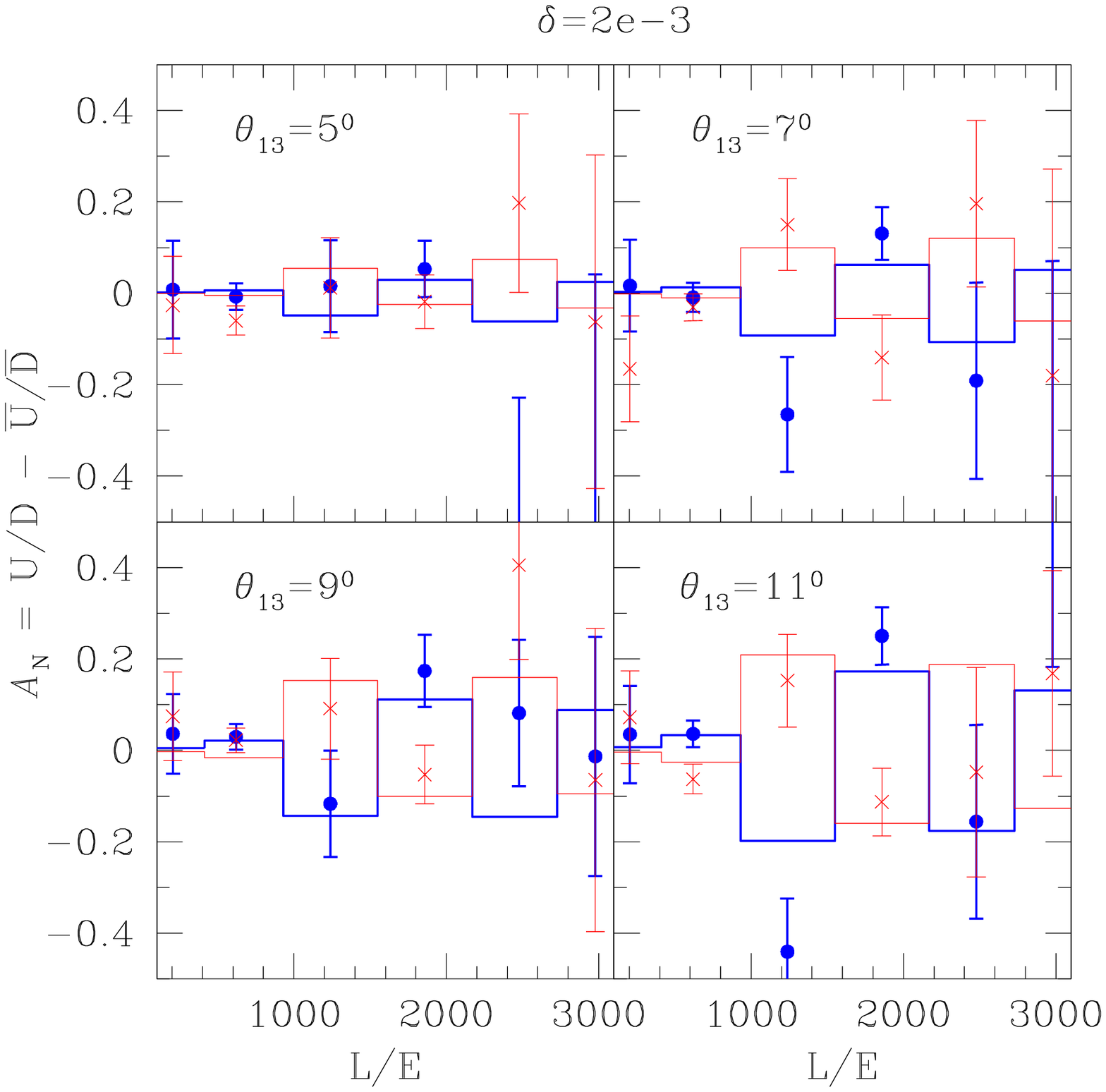}}
\caption{The same as Fig.~\ref{le1}
for $\vert \delta_{32} \vert = 2 \times 10^{-3}$ eV${}^2$.}
\label{le2} 
\end{figure}

\begin{figure}[htp]
\vskip 14truecm
{\includegraphics{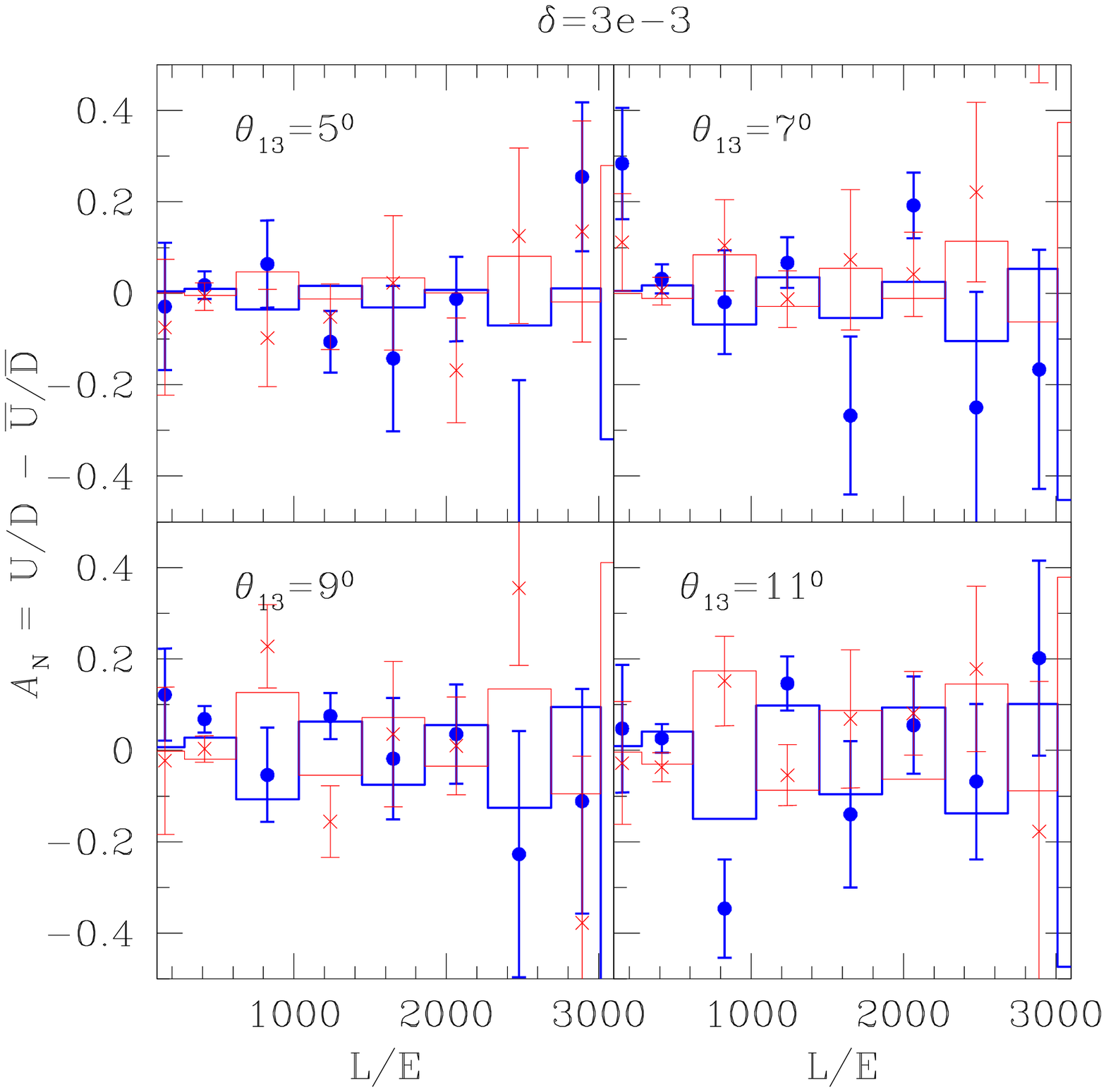}}
\caption{The same as Fig.~\ref{le1}
for $\vert \delta_{32} \vert = 3 \times 10^{-3}$ eV${}^2$.}
\label{le3} 
\end{figure}

We exhibit the results for $\vert \delta_{32} \vert = 1, 2, 3
\times 10^{-3}$ eV${}^2$ in Figs.~\ref{le1}, \ref{le2} and \ref{le3}
respectively. In each figure the events asymmetry ${\cal{A}}_N$ is plotted
as a function of $L/E$.  In all cases, a zenith angle cut of $\vert
\cos\theta \vert > 0.1$ was used to remove ``horizontal'' events, as
discussed earlier. The first data point corresponds to a region where no
matter effect is expected, that is, up to an $L/E$ smaller than the $n =
1$ value in eq.~\ref{bins} above.

The ``data'' points shown correspond to binning the actual events
generated using the {\sc nuance} Monte Carlo event generator in bins whose
widths follow the expression eq.~\ref{bins}. The ``dots'' and the
``crosses'' correspond to the direct and inverted hierarchy. The
theoretical asymmetry generated by integrating the differential event
rate over the bin width is plotted for comparison as a histogram. The
darker lines correspond to the direct hierarchy and lighter ones to the
inverted hierarchy. The deviation of the points from the value indicated
by the corresponding histogram is an indicator of fluctuations. The size
of the error-bar on the data reflects only the statistical error due to
the data size in that bin.

The error bars are larger for the inverted than for the direct hierarchy
in a given bin. This is because it is the anti-neutrino event rate that
is sensitive to matter effects in the case of inverted hierarchy, as
described in section 2. Since the anti-neutrino cross-sections are about
a factor of two smaller than the neutrino cross-section, the event rates
are smaller and the error bars correspondingly larger. Statistically
speaking, therefore, it is easier to detect direct than inverted
hierarchy.

Results are separately shown for $\theta_{13} = 5, 7, 9, 11$ degrees. 
It is easily seen that the greatest sensitivity to matter effects occurs
for a value of $\vert \delta_{32} \vert = 2 \times 10^{-3}$ eV${}^2$.
For a smaller $\vert \delta_{32}\vert $, the matter effect does not develop
significantly until a fairly large  $L/E$, where the event rates begin
to drop; for larger $\vert \delta_{32} \vert$, the larger number of bins
where the asymmetry changes sign leads to smaller statistics per bin.

Note that the first oscillation period (second data point) corresponds
to a minimum in the up-coming events rate. In all cases, the largest
sensitivity to the hierarchy is therefore in the region around the first
maximum in the up/down ratio, or the third bin. In fact, the survival
probability is sensitive to matter effects for energies roughly in the
range $2 \le E \le 10$ GeV. The cut of 4 GeV on the neutrino energy
is used to accentuate the asymmetry in the third bin.

A negative (positive) value of asymmetry in this bin clearly indicates
direct (inverted) mass hierarchy. It may be possible to improve this
result by binning over regions of $L/E$ taking into account modifications
due to matter of eq.~\ref{bins}; in fact, with real data, it may be also
be possible to have variable bin sizes, determined by maximising the
magnitude of the asymmetry in each bin. We shall not attempt this here.
We only wish to note that the third bin is automatically identified once
the magnitude of $\delta_{32}$ is determined. It may also be possible to
combine the results of several bins (with appropriate sign) to improve
the statistics.

Note that with an energy cut $E > 4$ GeV, the events are roughly
distributed in the proportion $2.5 : 2 : 1$ from deep inelastic
scattering (DIS), resonance (dominated by 1-pion processes) (RS),
and quasi-elastic (QE) charged current processes respectively. There
are roughly 3000 events in 480 kton-years, of which about 35\% are
up-going ones.

Finally, if a non-zero asymmetry is measured, this will yield both the
mass hierarchy as well as a lower bound on $\theta_{13}$ of at least
$\theta_{13} > 6^\circ$ ($\sin^22\theta_{13} > 0.04$). While the
amplitude of the asymmetry is proportional to $\sin\theta_{13}$,
limitations due to statistics make it unlikely that the value of
$\theta_{13}$ itself will be determined with any reasonable accuracy. One
needs to address this issue elsewhere.

\subsection{Asymmetries with finite detector resolution}

Inclusion of finite detector resolution reduces the sensitivity,
especially beyond the first oscillation minimum and maximum. A
straightforward way to include such effects is to smear the observed $L$
and $E$ and observe the impact of this. To this end, we define
Gaussian resolution functions for both $L$ and $E$:
\begin{eqnarray}
R_1(E', E) & \equiv & \by{1}{\sqrt{{2}{\pi}}\sigma_1} \exp \left[
\by{-(E-E')^2}{2 \sigma_1^2} \right]~; \\ \nonumber
R_2(L', L) & \equiv & \by{1}{\sqrt{{2}{\pi}}\sigma_2} \exp \left[
\by{-(L-L')^2}{2 \sigma_2^2} \right]~.
\label{resfns}
\end{eqnarray}
Hence, the event rate now includes the probability that a neutrino of
any possible $L'$ and $E'$ is detected in the detector with path-length
$L$ and energy $E$. We have
\begin{equation}
N^{\alpha,R}_{\rm bin}(x) = \int_{\rm bin}  \d x \int_{E_{\rm min}}
\by{\d E}{E} \int_0^\infty \d E' R_1(E', E) \int_0^\infty \d L' R_2 (L', L) 
\frac{\d^2 N^{\alpha}}{\d\ln E'\d x'}~,
\label{eventR}
\end{equation} 
where\footnote{The combination of Gaussian resolution functions for $L$
and $E$ integrates to a Lorentzian in their ratio provided the integration
over {\em all} variables is unrestricted. We need to specify individual
resolution functions here because of the cut on $E$.} $x' = L'/E'$.

We re-evaluate the event rates and the up/down asymmetries using this
equation, with $\sigma_1 = 0.15 E'$ and $\sigma_2 = 0.15 L'$. These are
realistic widths obtained by a {\sc geant} analysis of atmospheric
neutrino events by both the {\sc monolith} and the {\sc ical/ino}
collaborations; this gives a typical FWHM for their ratio $(L/E)$ to be
about $\sim 0.4 L/E$.

\begin{figure}[htp]
\vskip 10truecm
\includegraphics{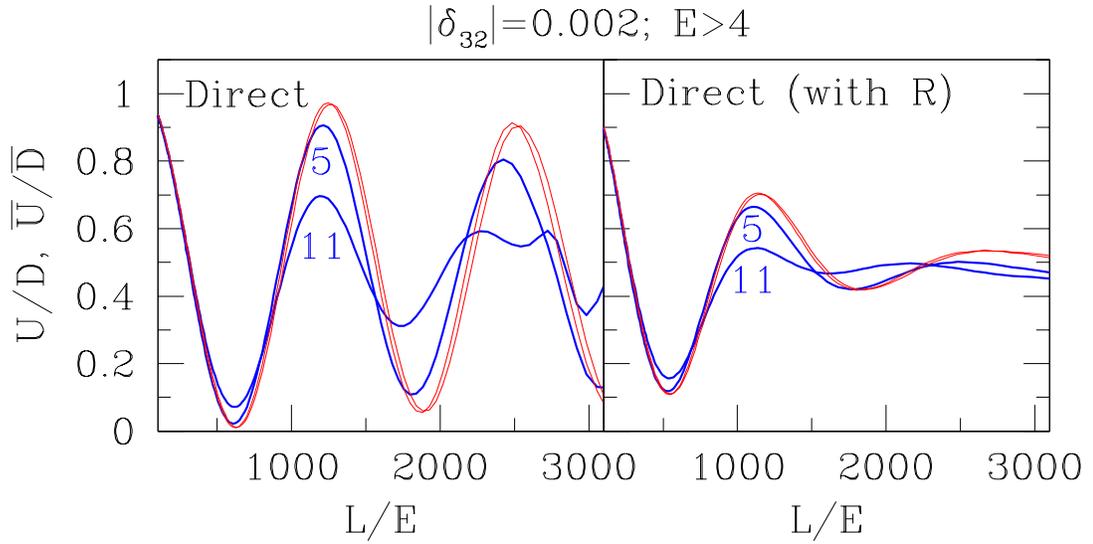}
\caption{A comparison of the up/down events ratio as a function of $L/E$
(for the direct hierarchy) with (right) and without (left) the inclusion
of Gaussian resolution functions for $\vert \delta_{32} \vert = 2 \times
10^{-3}$ eV$^2$ and $\theta_{13} = 5,11^\circ$. Results are very sensitive
to the widths of the Gaussian used. The results presented correspond
to a Gaussian width $\sigma = 15$\% for both $L$ and $E$. The thick
(blue) curves correspond to neutrinos and the thin (red) curves to
the anti-neutrinos; note that the latter are barely discernible from
each other.}

\label{le511R} 
\end{figure}

The up/down events ratio (for the case of the direct hierarchy) is shown
on the right side of Fig.~\ref{le511R} as a function of $L/E$ for
$\delta_{32} = 2 \times 10^{-3}$ eV$^2$. The results without including
resolution functions is shown in the left-hand panel, for comparison. As
before, a cut on the neutrino energy of $E > 4$ GeV is applied. With the
inclusion of the resolution functions, it is seen that the {\em amplitude}
of the oscillation is greatly damped. Furthermore, the neutrino rates
ratio (shown as thick blue lines) is damped much more than that for the
anti-neutrinos (the converse will be true for the inverted hierarchy).
Curves correspond to $\theta_{13} = 5, 11^\circ$. Clear oscillations
are visible only around the first minimum and maximum.

\begin{figure}[htp]
\vskip 14truecm
\includegraphics{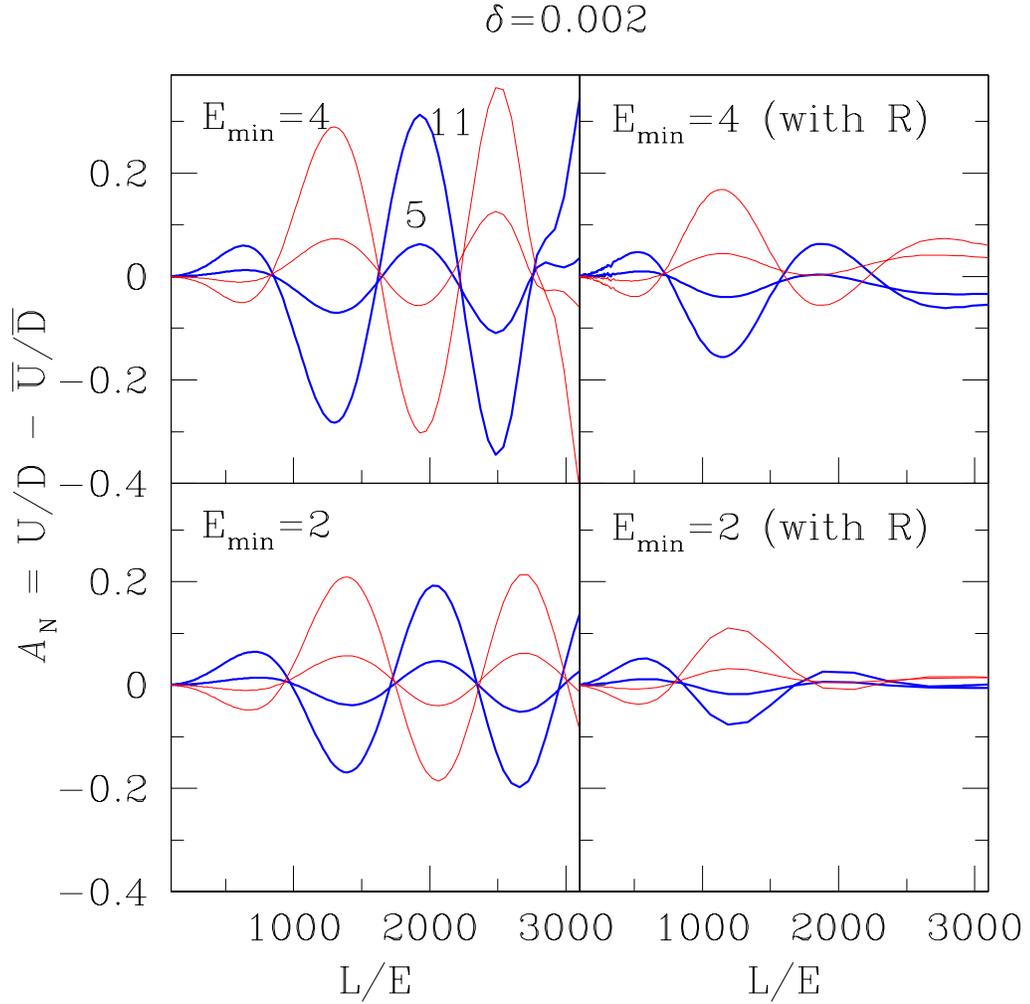}
\caption{A comparison of the difference asymmetry as in Fig.~\ref{ledif}
with (right) and without (left) inclusion of resolution functions for
$\delta \equiv \vert \delta_{32} \vert = 2 \times 10^{-3}$ eV$^2$ and
$\theta_{13} = 5,11^\circ$. The thick (blue) curves correspond to the
direct and the thin (red) curves to the inverted hierarchy. Two cases,
$E > 4$ and $E > 2$ GeV, are shown. It is clear that the higher energy cut
better emphasises the asymmetry and hence the matter effects; however, the
event rates are roughly halved in this case, for the same exposure times.}

\label{ledifR} 
\end{figure}

The resulting asymmetry, that is, the difference of the neutrino
and anti-neutrino events ratios, ${\cal{A}}_N$, is plotted in
Fig.~\ref{ledifR}. Again, the panels on the right correspond to the
rates calculation including resolution, while those on the left
correspond to rates without including resolution effects. There is
virtually no difference for small $L/E$. In general, however, the amplitude
of the asymmetry decreases although it is still largest in the second
envelope (corresponding to the third bin in the earlier discussions).

It is seen that the asymmetry is larger with a cut of $E_{\rm min} = 4$
GeV than with a cut of $E_{\rm min} = 2$ GeV while the {\em number
of events} is smaller by roughly half in the case of the larger energy
cut. In particular, by lowering the cut to $E_{\rm min} = 2$ GeV, not
only does the sample size effectively double, the contribution from
various processes changes to the proportion $1 : 1.5 : 1$ from $2.5 : 2 :
1$ for DIS, RS and QE processes.

Thus there are two conflicting requirements: to enhance the size of
the asymmetry, we need a larger value of $E_{\rm min}$, but to improve
statistics, we need a larger sample size. The final choice of $E_{\rm
min}$ should optimise these two requirements.

We note in passing that, for $E_{\rm min} = 4$ GeV, there is a fairly
large asymmetry for large $L/E$ as well; the statistical significance in
this bin is however small. In both cases, the asymmetry in the penultimate
envelope is almost vanishingly small; the oscillation length in this
case is also very small.

In short, it appears that a finite resolution function causes smearing
of the data across bins in such a way that the magnitude of the
asymmetry becomes significantly smaller, and even vanishes beyond the
third envelope/bin. We can improve the result by summing the asymmetries
over several envelopes. We define
\begin{equation}
{\cal{A}}_n^H = \sum_{i=1}^n (-1)^i {\cal {A}}_N^{H,i}~,
\label{ANH}
\end{equation}
where $H$ corresponds to a given hierarchy, direct, or inverted, $H=
D, I$. Typically, ${\cal{A}}_N^{D} \sim - {\cal{A}}_N^{I}$ in a bin,
and therefore the sum in eq.~\ref{ANH} is roughly equal and opposite for
the two hierarchies. Furthermore, the statistics are not very different
with either hierarchy. The significance of the result is thus estimated
by comparing the difference
\begin{equation}
\Delta {\cal{A}}_n = {\cal {A}}_n^D - {\cal {A}}_n^I~
\label{deltaAN}
\end{equation}
with the error $\sigma$ corresponding to a given exposure. In fact,
the ratio $\Delta{\cal{A}}_n/\sigma$ indicates the probability of the
observed asymmetry being assigned to the correct hierarchy.

The error $\sigma$ arises primarily from statistics. This is because the
asymmetry is defined as a difference of two ratios. Systematic errors
(due to uncertainties both in fluxes and cross-sections) in the ratio
are therefore very small and can be ignored; this was pointed out very
early on by the {\sc monolith} collaboration \cite{monolith}.

We list the values of $\Delta{\cal{A}}_n$ and the corresponding
statistical error $\sigma_n$ (we have chosen the larger of the errors
for the direct and inverted hierarchy) in Tables \ref{table15el},
\ref{table10el}, \ref{table05el} for different resolutions in $E$ and
$L$ of 15\%, 10\%, and 5\% respectively. In each table, results for
both $n = 2$ and $n = 3$ (that is, a sum over two or three envelopes),
corresponding to $\vert \delta_{32} \vert = 2.0 \times 10^{-3}$ eV$^2$
and $\theta_{13} = 7, 9, 11, 13^\circ$ are shown for different exposures
of 480, 800 and 1120 kton-years respectively.  The resulting confidence
level of the measurement is also indicated in the tables.

It is clear that the asymmetry and hence $\Delta {\cal{A}}_n$ increases
with $\theta_{13}$, {\it independent} of exposure. The effect of
increasing the exposure is to improve the errors, without changing the
central values of the asymmetry. This can be seen by comparing results
for the same $\theta_{13}$ in each table. Since the minimum exposure
of 480 kton-years is already large, and statistical errors improve
only as $\sqrt{N_{data}}$, it will be difficult to greatly enhance the
significance of the result by going to larger and larger exposures.

The effect of improving the resolution function (obtained by comparing
the same row in the tables \ref{table15el}, \ref{table10el}
and \ref{table05el} is two-fold. For a given exposure, improving
the resolution function (by which we mean the improving the Gaussian
widths of the resolutions in $E$ and $L$ as defined in eq.~\ref{resfns}
above) {\it increases the central value of the asymmetry} in each case.
It also results in marginally improved errors for $n=3$, that is, when
summing over all the three significant bins in the $L/E$ distribution,
which further enhances the significance of the result. For example, the
improvement upon increasing the exposure from 480 to 1120 kton-years at
15\% resolution is similar to that obtained by improving the resolution,
{\it at 480 kton-years}, from 15\% to 10\%.

Note also that the significance of the results for $\Delta {\cal{A}}_2$
is higher than for $\Delta {\cal{A}}_3$ for larger resolutions of 15\%;
this may imply that best results are obtained from single-bin
information when the resolutions are poorer. Also, it turns out that the
results are systematically better for a cut of $E_{min} = 4$ GeV rather
than for a cut of 2 GeV.

While resolutions with widths of 10\% may not be hard to achieve, it is
not clear whether it is possible to reach 5\%. One way is to exclude
the DIS events (perhaps by cutting out events with large hadronic
activity). The resulting data set is smaller by about 45\%; however,
it may be possible to resolve the energy and direction of the neutrino
much better since there are only quasi-elastic and resonance events left
in the sample. This needs to be studied in more detail.

In summary, it appears that, even for moderate to large exposures, it may
be difficult to achieve effects larger than $2\sigma$ or so, unless of
course Nature is kind and prefers large $\theta_{13}$.  For $\theta_{13}
\ge 9^\circ$, a significance of 99.3\% (nearly $3\sigma$) or more is
obtained with an exposure of 800 (1120) kton-years and resolutions in both
energy and angle of 5\% (10\%).  It is clear that a prior knowledge of
$\theta_{13}$ from some other experiment would greatly facilitate the
extraction of the mass hierarchy since the magnitude of the expected
asymmetry would then be known.

In short, the major features of the analysis without including
finite detector resolutions appear to hold, although with much reduced
significance. A prima facie case has been made for determining mass
hierarchy through observation of atmospheric neutrinos with {\sc ical}
detectors; however, the effect is small and is difficult to measure.
Indeed, it is clear that the asymmetry is very sensitive to smearing in
both $L$ and $E$. Our study indicates that an {\it improvement in the
detector resolution} has a stronger impact on the results than a
mere enhancement in size or the number of years of data-accumulation.
A detailed study involving actual reconstruction of muon and hadron
energies and muon direction is needed to determine the ultimate
sensitivity of such a detector to the issue of hierarchy.

\subsection{The Sum}

\begin{figure}[htp]
\vskip 14truecm
{\includegraphics{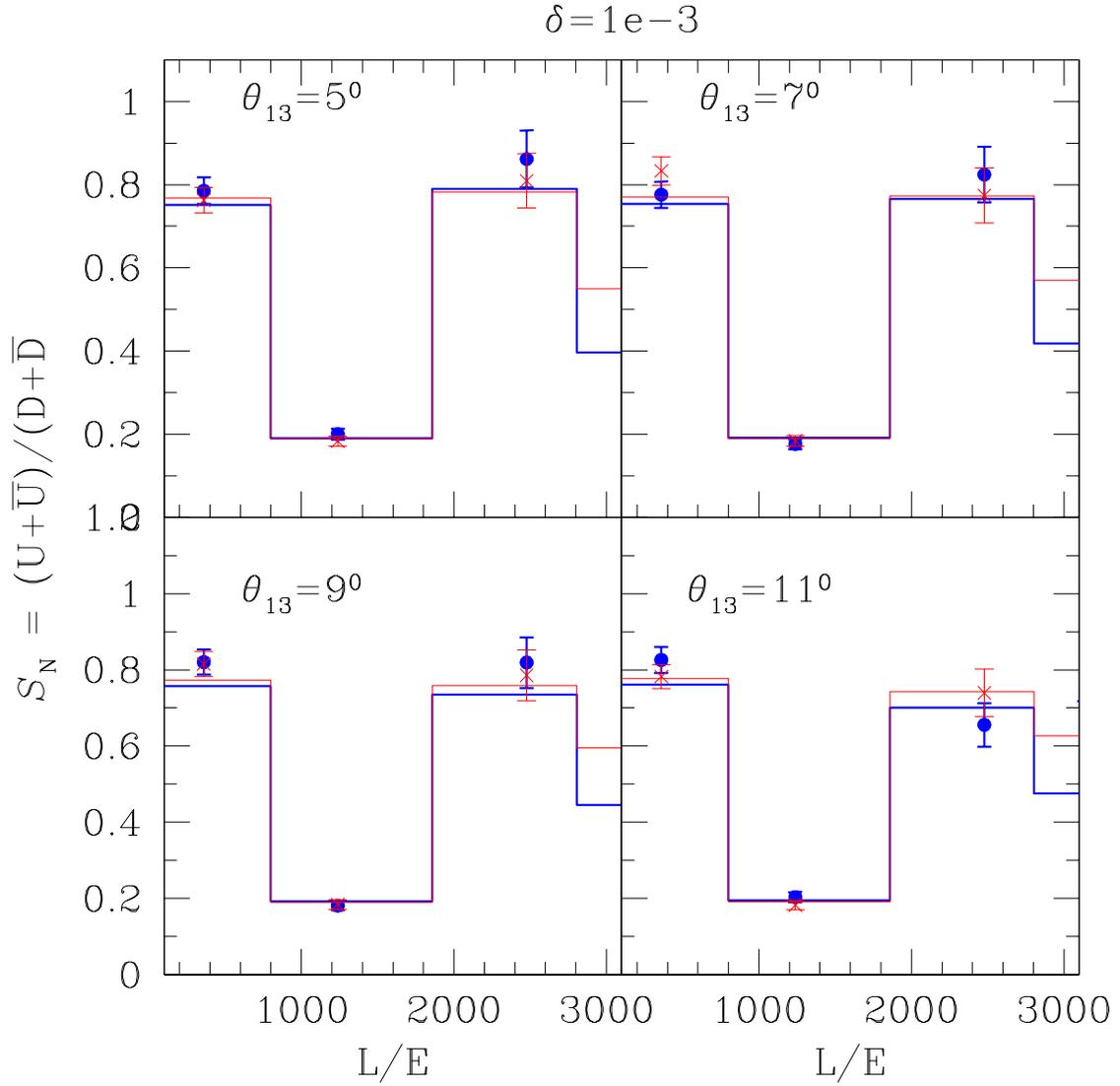}}
\caption{The up/down ratio ${\cal{S}}_N$ of the sum of neutrino and
anti-neutrino events integrated over $L/E$ bins shown with
statistical errors as a function of $L/E$. Results are shown for
$\theta_{13} = 5, 7, 9, 11$ degrees for $\vert \delta_{32} \vert =
1 \times 10^{-3}$ eV${}^2$.  For more details, see the caption of
Fig.~\ref{le1}.}

\label{lep1} 
\end{figure}

\begin{figure}[htp]
\vskip 14truecm
{\includegraphics{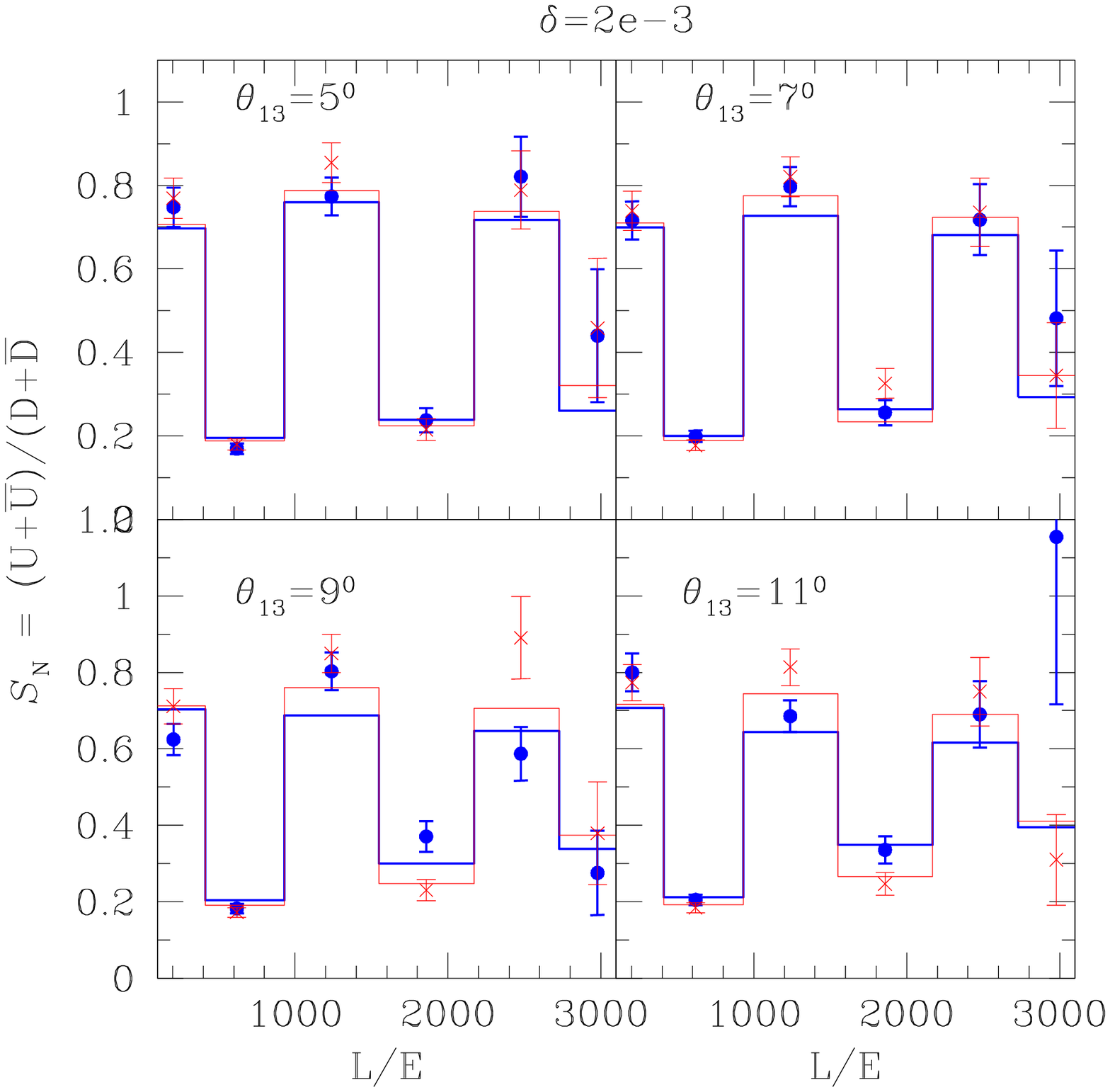}}
\caption{The same as Fig.~\ref{lep1}
for $\vert \delta_{32} \vert = 2 \times 10^{-3}$ eV${}^2$.}
\label{lep2} 
\end{figure}

\begin{figure}[htp]
\vskip 14truecm
\includegraphics{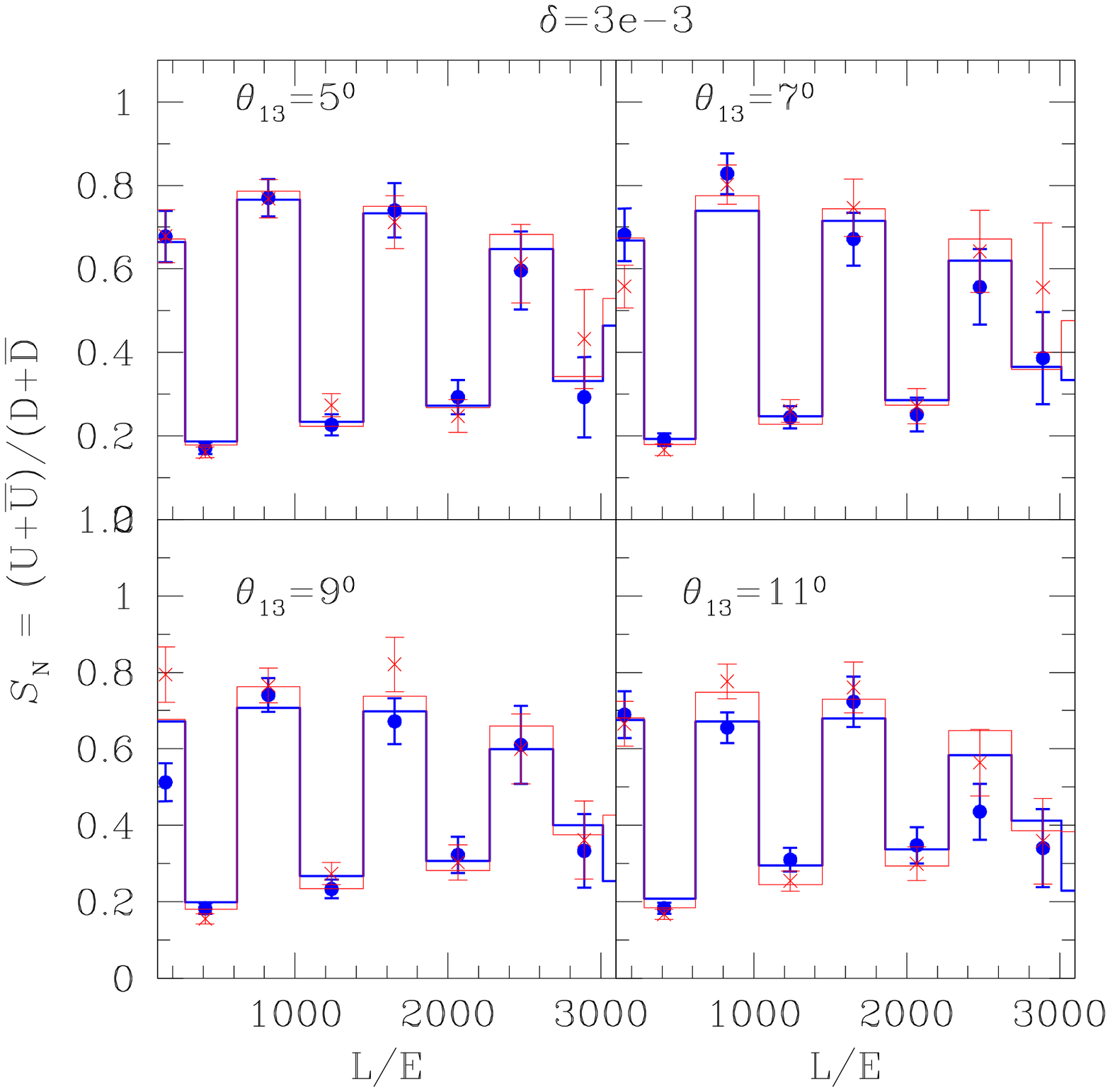}
\caption{The same as Fig.~\ref{lep1}
for $\vert \delta_{32} \vert = 3 \times 10^{-3}$ eV${}^2$.}
\label{lep3} 
\end{figure}

The results from the sum of neutrino and anti-neutrino events
${\cal{S}}_N$ are much less sensitive to matter effects, as
expected. Results (with same cuts as above) are shown in Figs.~\ref{lep1},
\ref{lep2} and \ref{lep3} for $\vert \delta_{32} \vert = 1, 2, 3 \times
10^{-3}$ eV${}^2$ respectively. For large values of $\theta_{13} >
7^\circ$, the amplitude of the sum averages to 1/2 much faster for the
direct hierarchy. The numerical study will need to be carefully done to
establish the significance of this result.

\section{Discussion}

We have discussed the possibility of obtaining information on the mixing
angle $\theta_{13}$ and the sign of $\delta_{32}$ using a magnetised iron
calorimeter collecting data on atmospheric muon neutrinos, capable of
charge identification. It appears that such a detector will be sensitive
to the neutrino mass hierarchy only if $\theta_{13} > 6^\circ$
($\sin^2 2 \theta_{13} > 0.04$).

We have defined a difference asymmetry in eq.~\ref{rateasym} as the
difference between the up/down rates of neutrinos and anti-neutrinos. By
taking the up/down ratio, the systematic errors are greatly reduced. In
particular, the uncertainty in the overall flux normalisation is
removed. Such a ratio also minimises the dependence on the cross-section,
which is substantially different for neutrinos and anti-neutrinos in
the GeV energy range of interest. As a consequence, the measurable
rate asymmetry is related to a theoretical flux-weighted probability
(difference) asymmetry given in eq.~\ref{asym}. Examination of the
theoretical asymmetry immediately suggests ways of binning the data in
order to maximise the asymmetry in each bin (in $L/E$). It turns out
that the {\em sign of the asymmetry} is sensitive to the neutrino mass
hierarchy (that is, the sign of $\delta_{32} = m_3^2 - m_2^2$) while
its magnitude is sensitive to the (13) mixing angle $\theta_{13}$.

We summarise our conclusions below.
\begin{itemize}

\item The limits on $|\delta_{32}|$ and $\sin 2\theta_{23}$ may be
reliably obtained using the location of the first minimum and the
amplitude of first oscillation of a plot of the ratio of up/down events
as a function of $L/E$. While no detailed discussion on this issue
has been presented here, it has been well-established elsewhere
\cite{monolith,ino}. 

The detector mass-exposure required for this purpose is about 150
kton-years. While charge identification is not crucial for this
step, presence of a magnetic field helps in obtaining better energy
resolution. This step however is {\it not dependent} on the value of
$\theta_{13}$ or the sign of $\delta_{32}$.

\item If $\theta_{13}$ is larger than $6^\circ$, then the pattern of
subsequent oscillations after the first dip can give definite indication
about the hierarchy through the difference asymmetry as discussed above.
It is possible to maximise the sensitivity to this asymmetry in fixed $L/E$
bins (improve statistical significance) by applying an energy cut,
$E_{\rm min} > 4$ GeV. An optimum $\vert\delta_{32}\vert$ dependent
bin-width can be determined that is {\em matter independent}. This is
easily accomplished as $\vert \delta_{32} \vert$ can be well determined
independently as discussed above.

For direct hierarchy the matter effects dominate through the neutrinos
which have a higher cross section and hence give more pronounced matter
effects. If the matter effects are small even when $\theta_{13}$ is large,
then it is a clear indication of inverted hierarchy since the dominant
neutrino events will have less impact due to matter.

\item Inclusion of finite detector resolution reduces the sensitivity,
especially beyond the first oscillation minimum and maximum in
$L/E$. A combination of enhanced exposure and improved resolution will
still enable a measurement of the mass hierarchy, provided $\theta_{13}$
is not too small, $\theta_{13} > 6^\circ$. The best scenario is
when $\theta_{13}$ is already established from other experiments to be
in the favourable range. Of course, in the absence of such a
measurement, non-observation of the asymmetry can put a lower bound on
$\theta_{13}$ while leaving open the issue of the hierarchy.

\item It is therefore important to maximise the sensitivity of the
detector to the range of $L/E$ where the matter effects are significant. 
It may be interesting to consider a non-horizontal detector geometry
in order to maximise the sensitivity to the relevant $L/E$ bins;
typically, these correspond to an average path-length $L$ around
7000 km. If neutrinos with such path-lengths are to be transverse
to the iron plates, a geometry with plates tilted at about $33^\circ$
to the horizontal may be preferred. This needs more careful study.

\end{itemize}

While many experiments (present and proposed) will be sensitive to the
(13) mixing angle $\theta_{13}$, it appears that charge identification
may be essential for settling the issue of the mass hierarchy. In
this context we note that reactor neutrino experiments are not
sensitive to matter effects and will be complementary to an {\sc ical}
detector. Short-baseline accelerator experiments may also not be as
sensitive to matter effects. Long-baseline experiments with neutrino
factory beams can surely settle this issue, and are sensitive to much
smaller values of $\theta_{13}$, but are expected to come on-line at
much later times.

Thus a study of atmospheric neutrinos with a magnetised iron
calorimeter detector can, in principle, provide fundamental information
on the parameters of the neutrino mixing matrix and the neutrino
mass-squared differences; in particular, it may resolve the question of
hierarchy in the neutrino masses. It is certain that large exposures
(more than 500 kton-years) will be required to obtain a result with a
significance of at least 90\% CL. In addition, good resolution in both
energy and angle discrimination is required. For larger $\theta_{13} \ge
9^\circ$, a significance of 99.3\% (nearly $3\sigma$) or more is obtained
with an exposure of 800 (1120) kton-years and resolutions in both energy
and angle of 5\% (10\%). Results for smaller $\theta_{13}$ remain at the
($2\sigma$) 95\% CL even in the best-case scenario. Hence the practical
issue of experimental feasibility seems much harder to resolve.

\vspace{0.2cm}

\noindent{\bf Acknowledgements}:
We are grateful to Dave Casper for making the {\sc nuance} software freely
available, and answering a long list of questions on its use.  We are
grateful to the members of the {\sc ino} group, and in particular to the
members of the {\sc ino} group at IMSc, K. Kar, H.S. Mani, G. Rajasekaran,
and Abdul Salam for many discussions and encouragement. We also thank G.
Rajasekaran for a careful and critical reading of the manuscript and
many suggestions for its improvement. We thank Nita Sinha and Probir Roy
for pointing out errors in some analytical formulae of Section IIB.

\newpage
\begin{table}[thp]
\begin{tabular}{|c|l|l|} \hline
 Exposure (kton-year) & $\Delta {\cal{A}}_2$ (CL \%) & $\Delta
 {\cal{A}}_3$ (CL \%) \\  \hline
\multicolumn{3}{|c|}{$\theta_{13} = 7^\circ$} \\
 480    & $0.166 \pm 0.161  \,~~~(1.0 \,\sigma,~ 68.3\%)$ &
         $0.167 \pm 0.230  \,~~~(0.7 \,\sigma,~ 51.6\%)$ \\
 800    & $0.166 \pm 0.125  \,~~~(1.3 \,\sigma,~ 80.6\%)$ &
         $0.167 \pm 0.178  \,~~~(0.9 \,\sigma,~ 63.2\%)$ \\
 1120  & $0.166 \pm  0.105 \,~~~(1.6 \,\sigma,~ 89.0\%)$ &
         $0.167 \pm 0.151  \,~~~(1.1 \,\sigma,~ 72.9\%)$ \\ \hline
\multicolumn{3}{|c|}{$\theta_{13} = 9^\circ$} \\
 480    & $0.257 \pm 0.159  \,~~~(1.6 \,\sigma,~ 89.0\%)$ &
         $0.280 \pm 0.230  \,~~~(1.2 \,\sigma,~ 77.0\%)$ \\
 800    & $0.257 \pm 0.123  \,~~~(2.1 \,\sigma,~ 96.4\%)$ &
         $0.280 \pm 0.178  \,~~~(1.6 \,\sigma,~ 89.0\%)$ \\
 1120  & $0.257 \pm  0.104 \,~~~(2.5 \,\sigma,~ 98.8\%)$ &
         $0.280 \pm 0.150  \,~~~(1.9 \,\sigma,~ 94.3\%)$ \\ \hline
\multicolumn{3}{|c|}{$\theta_{13} = 11^\circ$} \\
 480    & $0.357 \pm 0.157  \,~~~(2.3 \,\sigma,~ 97.9\%)$ &
         $0.415 \pm 0.230  \,~~~(1.8 \,\sigma,~ 92.8\%)$ \\
 800    & $0.357 \pm 0.122  \,~~~(2.9 \,\sigma,~ 99.6\%)$ &
         $0.415 \pm 0.178  \,~~~(2.3 \,\sigma,~ 97.9\%)$ \\
 1120  & $0.357 \pm  0.103 \,~~~(3.5 \,\sigma,~ 99.95\%)$ &
         $0.415 \pm 0.150  \,~~~(2.8 \,\sigma,~ 99.6\%)$ \\ \hline
\multicolumn{3}{|c|}{$\theta_{13} = 13^\circ$} \\
 480    & $0.459 \pm 0.155  \,~~~(3.0 \,\sigma,~ 99.7\%)$ &
         $0.560 \pm 0.230  \,~~~(2.4 \,\sigma,~ 99.4\%)$ \\
 800    & $0.459 \pm 0.120  \,~~~(3.8 \,\sigma,~ 99.98\%)$ &
         $0.560 \pm 0.178  \,~~~(3.1 \,\sigma,~ 99.8\%)$ \\
 1120    & $0.459 \pm 0.102  \,~~~(4.5 \,\sigma,~ 99.999\%)$ &
         $0.560 \pm 0.151  \,~~~(3.7 \,\sigma,~ 99.98\%)$ \\
\end{tabular}
\caption{Significance of the asymmetry for different exposures in
kton-years for resolution widths $\sigma_{1,2}$ in $E$ and $L$
(see eq.~\ref{resfns}) of 15\%. Asymmetries are shown for typical values
of $\theta_{13} = 7, 9, 11, 13^\circ$; the {\sc chooz} bound limits this
angle to $\theta_{13} \le 14.9^\circ$.}
\label{table15el}
\end{table}

\begin{table}[thp]
\begin{tabular}{|c|l|l|} \hline
Exposure (kton-year) & $\Delta {\cal{A}}_2$ (CL \%) & $\Delta
 {\cal{A}}_3$ (CL \%) \\  \hline
\multicolumn{3}{|c|}{$\theta_{13} = 7^\circ$} \\
 480    & $0.185 \pm 0.167  \,~~~(1.1 \,\sigma,~ 72.9\%)$ &
         $0.232 \pm 0.220  \,~~~(1.1 \,\sigma,~ 72.9\%)$ \\
 800    & $0.185 \pm 0.130  \,~~~(1.4 \,\sigma,~ 83.8\%)$ &
         $0.232 \pm 0.170  \,~~~(1.4 \,\sigma,~ 83.8\%)$ \\
 1120  & $0.185 \pm  0.110 \,~~~(1.7 \,\sigma,~ 91.1\%)$ &
         $0.232 \pm 0.144  \,~~~(1.6 \,\sigma,~ 89.0\%)$ \\ \hline
\multicolumn{3}{|c|}{$\theta_{13} = 9^\circ$} \\
 480    & $0.287 \pm 0.165  \,~~~(1.7 \,\sigma,~ 91.1\%)$ &
         $0.384 \pm 0.220  \,~~~(1.8 \,\sigma,~ 92.8\%)$ \\
 800    & $0.287 \pm 0.128  \,~~~(2.2 \,\sigma,~ 97.2\%)$ &
         $0.384 \pm 0.170  \,~~~(2.3 \,\sigma,~ 97.9\%)$ \\
 1120  & $0.287 \pm  0.108 \,~~~(2.7 \,\sigma,~ 99.3\%)$ &
         $0.384 \pm 0.144  \,~~~(2.7 \,\sigma,~ 99.3\%)$ \\ \hline
\multicolumn{3}{|c|}{$\theta_{13} = 11^\circ$} \\
 480    & $0.399 \pm 0.163  \,~~~(2.4 \,\sigma,~ 98.4\%)$ &
         $0.565 \pm 0.221  \,~~~(2.6 \,\sigma,~ 99.1\%)$ \\
 800    & $0.399 \pm 0.126  \,~~~(3.2 \,\sigma,~ 99.9\%)$ &
         $0.565 \pm 0.171  \,~~~(3.3 \,\sigma,~ 99.9\%)$ \\
 1120  & $0.399 \pm 0.107 \,~~~(3.7 \,\sigma,~ 99.98\%)$ &
         $0.565 \pm 0.144  \,~~~(3.9 \,\sigma,~ 99.99\%)$ \\ \hline
\multicolumn{3}{|c|}{$\theta_{13} = 13^\circ$} \\
 480    & $0.515 \pm 0.161  \,~~~(3.2 \,\sigma,~ 99.9\%)$ &
         $0.758 \pm 0.223  \,~~~(3.4 \,\sigma,~ 99.9\%)$ \\
 800    & $0.515 \pm 0.125  \,~~~(4.1 \,\sigma,~ 99.996\%)$ &
         $0.758 \pm 0.173  \,~~~(4.4 \,\sigma,~ 99.999\%)$ \\
 1120    & $0.515 \pm 0.102  \,~~~(4.9 \,\sigma,~ 99.9999\%)$ &
         $0.758 \pm 0.146  \,~~~(5.2 \,\sigma,~ 100\%)$ \\
\end{tabular}
\caption{Same as Table \ref{table15el} for
resolution widths $\sigma_{1,2}$ in $E$ and $L$
(see eq.~\ref{resfns}) of 10\%.}
\label{table10el}
\end{table}

\begin{table}[thp]
\begin{tabular}{|c|l|l|} \hline
Exposure (kton-year) & $\Delta {\cal{A}}_2$ (CL \%) & $\Delta
 {\cal{A}}_3$ (CL \%) \\  \hline
\multicolumn{3}{|c|}{$\theta_{13} = 7^\circ$} \\
 480    & $0.198 \pm 0.172  \,~~~(1.1 \,\sigma,~ 72.9\%)$ &
         $0.297 \pm 0.209  \,~~~(1.4 \,\sigma,~ 83.8\%)$ \\ 
 800    & $0.198 \pm 0.133  \,~~~(1.5 \,\sigma,~ 86.6\%)$ &
         $0.297 \pm 0.162  \,~~~(1.8 \,\sigma,~ 92.8\%)$ \\
 1120  & $0.198 \pm  0.112 \,~~~(1.8 \,\sigma,~ 92.8\%)$ &
         $0.297 \pm 0.137  \,~~~(2.2 \,\sigma,~ 97.2\%)$ \\ \hline
\multicolumn{3}{|c|}{$\theta_{13} = 9^\circ$} \\
 480    & $0.308 \pm 0.170  \,~~~(1.8 \,\sigma,~ 92.8\%)$ &
         $0.489 \pm 0.209  \,~~~(2.3 \,\sigma,~ 97.9\%)$ \\
 800    & $0.308 \pm 0.132  \,~~~(2.3 \,\sigma,~ 97.9\%)$ &
         $0.489 \pm 0.162  \,~~~(3.0 \,\sigma,~ 99.7\%)$ \\
 1120  & $0.308 \pm 0.111  \,~~~(2.8 \,\sigma,~ 99.6\%)$ &
         $0.489 \pm 0.137  \,~~~(3.6 \,\sigma,~ 99.97\%)$ \\ \hline
\multicolumn{3}{|c|}{$\theta_{13} = 11^\circ$} \\
 480    & $0.429 \pm 0.168  \,~~~(2.5 \,\sigma,~ 98.8\%)$ &
         $0.717 \pm 0.213  \,~~~(3.4 \,\sigma,~ 99.9\%)$ \\
 800    & $0.429 \pm 0.130  \,~~~(3.3 \,\sigma,~ 99.9\%)$ &
         $0.717 \pm 0.165  \,~~~(4.3 \,\sigma,~ 99.9983\%)$ \\
 1120  & $0.429 \pm 0.110  \,~~~(3.9 \,\sigma,~ 99.99\%)$ &
         $0.717 \pm 0.139  \,~~~(5.1 \,\sigma,~ 100\%)$ \\ \hline
\multicolumn{3}{|c|}{$\theta_{13} = 13^\circ$} \\
 480    & $0.554 \pm 0.166  \,~~~(3.3 \,\sigma,~ 99.9\%)$ &
         $0.960 \pm 0.218  \,~~~(4.4 \,\sigma,~ 99.999\%)$ \\
 800    & $0.554 \pm 0.128  \,~~~(4.3 \,\sigma,~ 99.998\%)$ &
         $0.960 \pm 0.169  \,~~~(5.7 \,\sigma,~ 100\%)$ \\
 1120    & $0.554 \pm 0.109  \,~~~(5.1 \,\sigma,~ 100\%)$ &
         $0.960 \pm 0.143  \,~~~(6.7 \,\sigma,~ 100\%)$ \\
\end{tabular}
\caption{Same as Table \ref{table15el} for
resolution widths $\sigma_{1,2}$ in $E$ and $L$
(see eq.~\ref{resfns}) of 5\%.}
\label{table05el}
\end{table}

\newpage

\newpage
\newpage

\newpage

\newpage

\newpage

\newpage

\newpage

\newpage

\newpage

\newpage

\newpage

\newpage

\end{document}